\documentclass[iop,apj,tighten]{emulateapj}
\usepackage{natbib}
\usepackage{amsmath,amstext}
\usepackage{graphicx} 
\usepackage{comment}
\usepackage{multirow}

\usepackage{amssymb}

\begin{document}
\title[]{On the Chemical and Kinematic Consistency Between N-rich Metal-poor Field Stars and  Enriched Populations in Globular Clusters
}
\author{Baitian Tang\altaffilmark{1}}
\author{J. G. Fern\'andez-Trincado\altaffilmark{2}}
\author{Chao Liu\altaffilmark{3,4}}
\author{Jincheng Yu\altaffilmark{1}}
\author{Hongliang Yan\altaffilmark{3,4}}
\author{Qi Gao\altaffilmark{3,4}}
\author{Jianrong Shi\altaffilmark{3,4}}
\author{Douglas Geisler\altaffilmark{5,6,7}}
\affil{$^1$School of Physics and Astronomy, Sun Yat-sen University, Zhuhai 519082, China; tangbt@mail.sysu.edu.cn}
\affil{$^{2}$Instituto de Astronom\'ia y Ciencias Planetarias, Universidad de Atacama, Copayapu 485, Copiap\'o, Chile.}
\affil{$^3$Key Lab of Optical Astronomy, National Astronomical Observatories, Chinese Academy of Sciences, Beijing 100101, China}
\affil{$^4$School of Astronomy and Space Science, University of Chinese Academy of Sciences, Beijing 100049, China}
\affil{$^5$Departamento de Astronom\'{i}a, Casilla 160-C, Universidad de Concepci\'{o}n, Concepci\'{o}n, Chile}
\affil{$^6$Instituto de Investigaci\'on Multidisciplinario en Ciencia y Tecnolog\'ia, Universidad de La
Serena. Avenida Ra\'ul Bitr\'an S/N, La Serena, Chile}
\affil{$^7$Departamento de Astronom\'ia, Facultad de Ciencias, Universidad de La Serena. Av.
Juan Cisternas 1200, La Serena, Chile}

\begin{abstract}
Interesting chemically peculiar field stars may reflect their stellar evolution history and their possible origin in a different environment from where they are found now, which is one of the most important research fields in Galactic archaeology.  To explore this further, we have used the CN-CH bands around 4000 \AA~to identify N-rich metal-poor field stars in LAMOST DR3. Here we expand our N-rich metal-poor field star sample to $\sim 100$ stars in LAMOST DR5, where 53 of them are newly found in this work. We investigate light elements of the common stars between our sample and APOGEE DR14. While Mg, Al, and Si abundances generally agree with the hypothesis that N-rich metal-poor field stars come from enriched populations in globular clusters, it is still inconclusive for C, N, and O. After integrating the orbits of our N-rich field stars and a control sample of normal metal-poor field stars, we find that N-rich field stars have different orbital parameter distributions compared to the control sample, specifically, apocentric distances, maximum vertical amplitude (Zmax), orbital energy, and z direction angular momentum (Lz). The orbital parameters of N-rich field stars indicate that most of them are inner-halo stars. The kinematics of N-rich field stars support their possible GC origin. The spatial and velocity distributions of our bona fide N-rich field star sample are important observational evidence to constrain simulations of the origin of these interesting objects.

\end{abstract}

\maketitle

\section{Introduction}
\label{sect:intro}

With the release of Gaia DR2 \citep{Brown2018,Katz2018}, proper motions of billions of stars are now available to the astronomical community. Combining with radial velocities from large spectroscopic surveys, like Sloan Digital Sky Survey (SDSS, \citealt{Eisenstein2011, Blanton2017}), Gaia-ESO survey \citep{Gilmore2012, Randich2013}, and LAMOST Galactic spectroscopic survey \citep{Deng2012,Zhao2012}, the wealthy 6D information of billions of stars have challenged and even overthrown many aspects of our understanding of the Milky Way (MW). The discovery of snail shells in the phase space distribution of MW disk stars \citep{Antoja2018} has inspired the debates about their origin: whether they are generated by the passage of a dwarf galaxy (probably Sagittarius dwarf galaxy) through the MW disk \citep{Binney2018}, or they are the echo of the MW bar buckling \citep{Khoperskov2019}. Meanwhile, major accretion events begin to unveil themselves under the inspection of stellar distribution in various energy-momentum space \citep[e.g.,][]{Myeong2019}. These major accretion events injected most of the materials from the progenitor dwarf galaxies into our MW, including globular clusters (GCs). As GCs are one of the oldest objects in our Galaxy, identifying and studying accreted GCs help us to trace back the accretion history of our Galaxy. Though details of the major accretion events, e.g., the number of accretion events and the GCs associated with each event, are still under debate \citep[e.g.,][]{Helmi2018, Myeong2019, Massari2019}, it is widely accepted that a substantial number of halo stars and GCs were accereted \citep[e.g.,][]{Ostdiek2019}. Along the same line, more and more substructures, e.g., stellar streams, are identified inside the MW \citep[e.g.,][]{Malhan2018, Ibata2019b}. An increasing number of stellar streams are suggested to be related to the debris of (inner-halo) GCs \citep[e.g.,][]{Ibata2019a}. Besides these GC destruction events under the influence of Galactic potential, dynamical relaxation of GCs \citep[e.g.,][]{Weinberg1994, Vesperini1997} also eject member stars into the field. It would be of great interest to find such GC-ejected stars, in order to estimate the mass loss from GCs for a better understanding of our MW formation and evolution. To help achieve this goal, another characteristic of GCs is very helpful.

Most GCs are now found to host multiple populations (MPs) through photometry and spectroscopy \citep[e.g.,][]{Piotto2015, Milone2015, Carretta2010b, Meszaros2015,Tang2017,Tang2018}. Chemical abundances from spectroscopic data suggest that GCs have a group of so-called `second generation' (SG) stars with enhanced N, Na, (sometimes He and Al), but depleted C, O, (sometimes Mg). These kind of stars presumably are only formed in the dense environments of GCs. Therefore, identifying field stars with SG-like chemical pattern is a feasible way to find a link between field stars and GC ejection/dissolution. Thanks to large spectroscopic surveys, the search for these chemically peculiar stars is becoming more efficient. Using high spectral resolution surveys, multiple elements, like C, N, O, Na, Mg, and Al, can be measured, depending on the wavelength range and signal-to-noise ratio (S/N). Toward this, Apache Point Observatory Galactic Evolution Experiment \citep[APOGEE, ][]{Majewski2017} and Gaia-ESO survey have led to the discovery of a large group of N-rich field stars \citep{Lind2015, Martell2016, Schiavon2017chem, Fernandez-Trincado2016, Fernandez-Trincado2017a, FT2019c}. While high resolution spectra gives more elements for detailed investigation of their chemical history, 
low resolution spectra can supposedly extend the search for N-rich field stars toward fainter and more numerous samples \citep{Martell2010,Martell2011,Koch2019}. Simultaneously observing 4000 stars with fibers makes LAMOST an unprecedented machine in collecting low resolution stellar spectra. Using the CN-CH band features around 4000\AA, we have identified $\sim40$ N-rich field stars\footnote{Also called CN-strong CH-normal stars in Paper I.} in LAMOST DR3 \citep[][hereafter Paper I]{Tang2019}. The derived N-abundances of these stars are clearly higher than that of the metal-poor field stars, indicating that (1) our sample is a bona fide sample of N-rich field stars; (2) the classical extra-mixing theory may not work for these stars. Moreover, a substantial fraction of retrograding N-rich field stars suggest that some N-rich field stars may be accreted. In this work, we expand our sample to $\sim 100$ N-rich field stars in LAMOST DR5 (Section \ref{sect:data}), making it more robust for drawing statistical conclusions, especially for the GC origin of these field stars. We put forward a detailed analysis of high-resolution chemical abundances and kinematics (Section \ref{sect:chem} and \ref{sect:orb}) to discuss the origins of these N-rich field stars (Section \ref{sect:dis}). As the second paper of this series, we will also call the present work Paper II.

\section{Sample Selection}
\label{sect:data}

\begin{figure}
\centering
\includegraphics [width=0.45\textwidth]{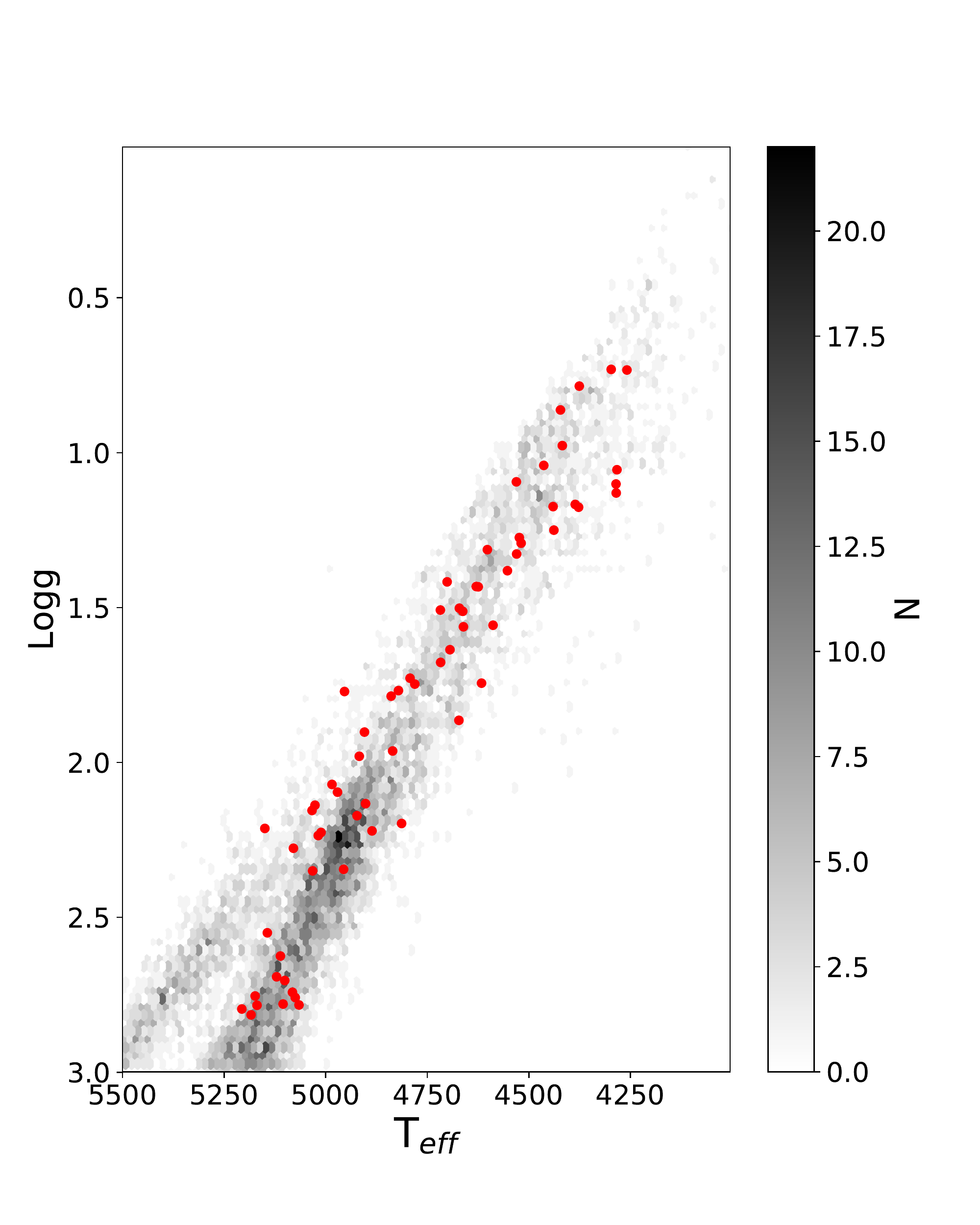} 
\caption{T$_{\rm eff}$ vs. $\log g$ plot. The gray scale map indicates the number density of metal-poor field stars that we selected as parent sample. The N-rich field stars are shown as red dots. 
}\label{fig:tlogg}
\end{figure}

\begin{figure*}
\centering
\includegraphics [width=1.05\textwidth]{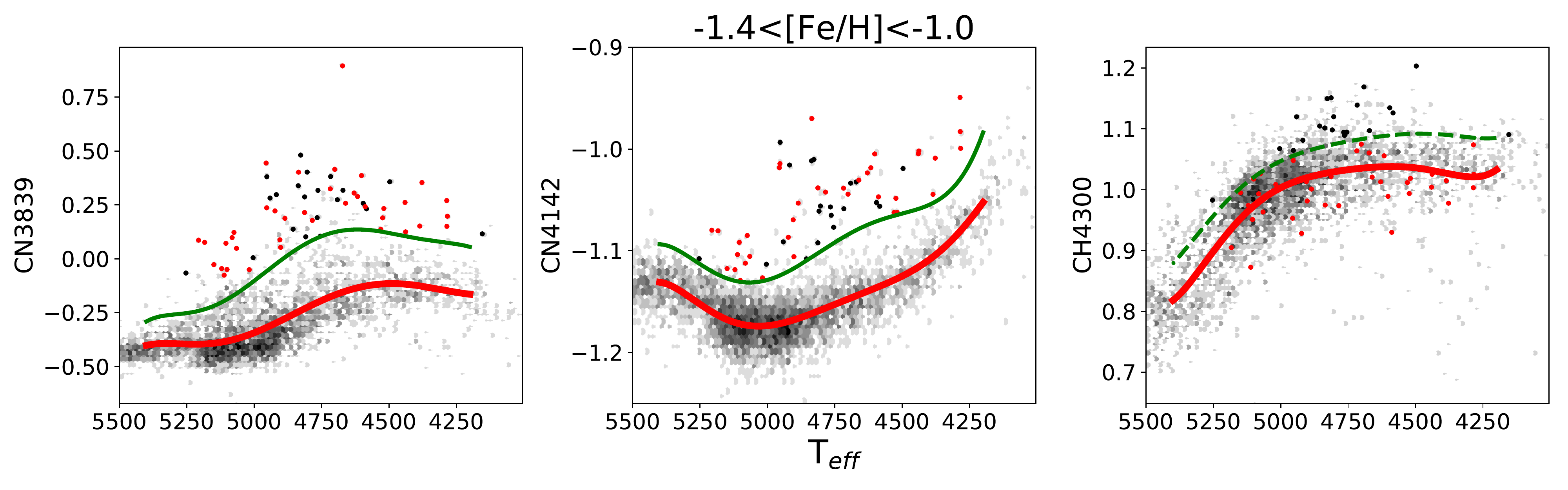} 
\includegraphics [width=1.05\textwidth]{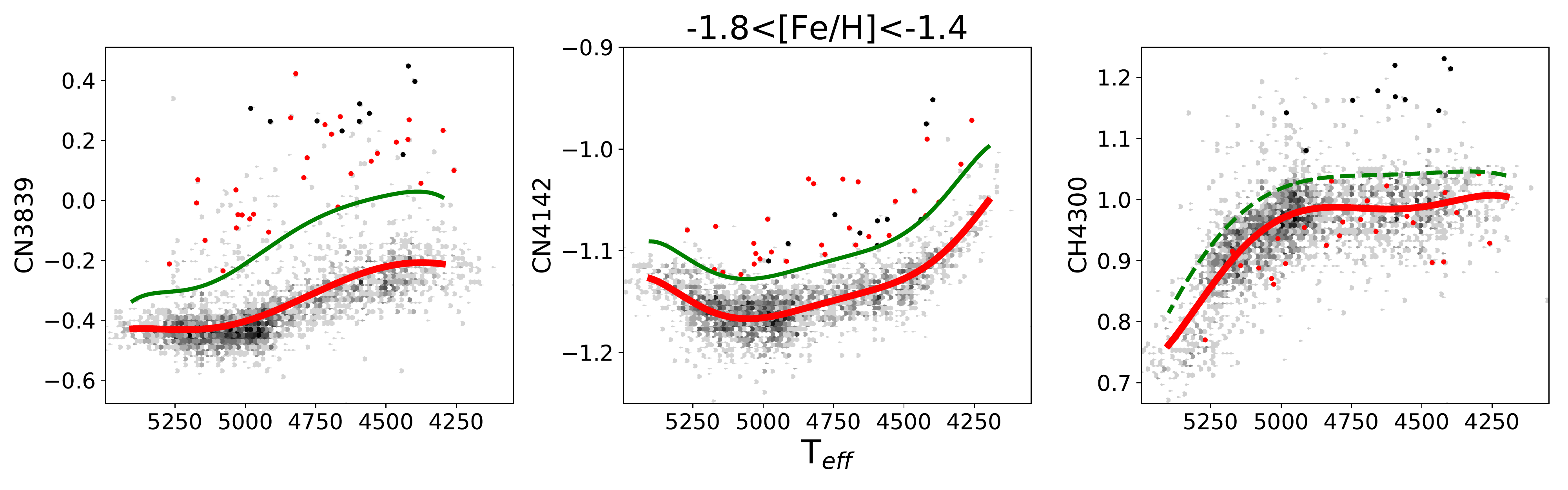} 
\caption{Spectral indices as a function of T$_{\rm eff}$. The gray scale map indicates the number density of metal-poor field stars. The upper panels show the more metal-rich sample ($-1.4<[$Fe/H$]<-1.0$), while the lower panels show the more metal-poor sample ($-1.4<[$Fe/H$]<-1.8$). 
Red lines are sixth-order polynomials of the mean spectral indices at a step of 100 K.  Green solid lines are sixth-order polynomials of the mean spectral indices plus 2 times standard deviations at a step of 100 K. Similarly, green dashed lines are $\rm mean+1.0\times std$. The black dots are CH-strong CN-strong (C-rich) stars, while the red dots are CH-normal CN-strong (N-rich) stars. See text for more details.
}\label{fig:CNCH}
\end{figure*}

We first select metal-poor RGB field stars according to the same criteria as described in Paper I, which is:
\begin{enumerate}
\item $4000<T_{\rm eff}<5500$ K
\item $\log$ g $< 3.0$
\item $-1.8<[$Fe/H$]<-1.0$
\item SNR$_u>5.0$
\end{enumerate} 
The stellar parameters are derived by the LAMOST Stellar Parameter pipeline \citep[LASP,][]{Wu2011, Luo2015}, with typical uncertainties of 100 K for T$_{\rm eff}$, 0.25 dex for $\log g$, and 0.1 dex for [Fe/H], respectively.

We first check for duplication in the sample and keep the observation with the highest SNR for a given star if there were multiple spectra. Then we remove all the stars that have been selected as the parent sample of Paper I. We also check for globular cluster membership using the membership identification method described in \citet{Tang2017}. Four possible members of M3 and M10 are excluded. Note that open cluster membership is unlikely, since all open clusters are more metal rich than [Fe/H$]=-1.0$. Thus, our final parent sample consists of 6592 metal-poor field red giants. The T$_{\rm eff}$ vs. $\log g$ plot is shown in Figure \ref{fig:tlogg}.

Next, we measure the spectral indices of our sample stars. Here we use the definition of CN3839, CN4142, and CH4300 indices from \citet{Harbeck2003}. When correcting for radial velocities (RVs) in the LAMOST spectra, we consider the systematic RV shift ($\sim 5$ km/s) reported by \citet{Schonrich2017}. 
The selection of N-rich field stars were described in Paper I, here we briefly summarize the selection method. Since CN and CH spectral indices are affected by stellar parameters, especially T$_{\rm eff}$, we plot both CN spectral indices as a function of T$_{\rm eff}$ to first select CN-strong stars that are 2$\sigma$ above the mean spectral indices of the parent sample stars at the same T$_{\rm eff}$ (Figure \ref{fig:CNCH} green solid lines). Only stars with both strong CN3839 and strong CN4142 are considered as CN-strong stars. Using both CN spectral indices is meant to reduce false positives in the CN-strong star sample. The CN-strong stars are further divided into CH-strong stars and CH-normal stars based on whether their CH spectral indices are above the one $\sigma$ line (Figure \ref{fig:CNCH} green dashed lines). As we found in Paper I and later show again in this paper, the CN-strong CH-normal stars are in fact N-rich stars, so we call them N-rich stars in the following text. One improvement that we carry out in this paper is that we subdivide the sample into a more metal-rich sample ($-1.4<[$Fe/H$]<-1.0$) and a more metal-poor sample ($-1.8<[$Fe/H$]<-1.4$), in order to further minimize the metallicity effect on spectral indices. 
We did not attempt to select CN-strong stars based on spectral index vs. absolute magnitude plots, because using absolute magnitude may introduce errors from the distance determination (Section \ref{sect:orb}). The uncertainties of spectral indices propagated from the typical uncertainties in LAMOST-derived stellar parameters \citep{Luo2015} are estimated using stellar atmosphere models. The estimated uncertainties of spectral indices are listed in Table \ref{tab:unc}. The uncertainties are almost negligible compared to the measurements (Table \ref{tab:cnch}).

Using this method, we select 67 N-rich stars from the parent sample. After visual examination of these 67 stars, we further exclude two stars with strong nebula emission lines. Thus we are left with 65 stars as the final sample of this paper (Table \ref{tab:nrich}). As a result, the N-rich field stars constitute about 1\% of our parent sample. 
If only CN3839 is used as the discriminator to separate CN-strong stars from CN-normal stars, the number of CN-strong CH-normal (N-rich) stars would increase to 160, and the total percentage of these stars in our sample would increase to 2.4\%. This percentage agrees with that of \citet{Martell2010, Martell2011} and \citet{Koch2019}.

\begin{table*}
\caption{Uncertainties of Spectral Indices.}              
\label{tab:unc}      
\centering                                      
\begin{tabular}{c | c c c c | c c c c }         
\hline\hline  
  &\multicolumn{4}{c|}{metal-rich sample}&\multicolumn{4}{c}{metal-poor sample}\\
Index &  $\sigma_{\Delta Teff}$ & $\sigma_{\Delta log(g)}$ & $\sigma_{\Delta [Fe/H]}$ & $\sigma_{total}$ &$\sigma_{\Delta Teff}$ & $\sigma_{\Delta log(g)}$ & $\sigma_{\Delta [Fe/H]}$ & $\sigma_{total}$\\
\hline 
CN3839  & 0.020 & 0.003 & 0.015 & 0.025 & 0.024 & 0.000  & 0.013  & 0.028\\ 
CN4142  & 0.002 & 0.003 & 0.002 & 0.005 & 0.007 & 0.006  & 0.003  & 0.009\\ 
CH4300  & 0.013 & 0.000 & 0.008 & 0.015 & 0.019 & 0.006  & 0.008  & 0.021\\ 
\hline                                             
\end{tabular}

\end{table*}

\begin{table*}
\caption{Spectral Indices of N-rich Field Stars Found In This Paper.}              
\label{tab:cnch}      
\centering                                      
\begin{tabular}{c c c c c c c c c }         
\hline\hline  
\# &  RA &  DEC & CN3839 & CN4142 & CH4300 & $\delta$CN3839 & $\delta$CH4300   & Note$^a$\\
\hline 
1  & 13.155076 & 37.698769 & 0.23 & -1.06 & 1.02 & 0.35  & -0.02  & MR\\ 
2  & 245.556824 & 2.385621 & 0.19 & -1.05 & 1.00 & 0.47  & -0.02  & MR\\ 
3  & 268.058059 & 26.255920 & 0.05 & -1.11 & 0.96 & 0.42  & -0.02  & MR\\ 
..  &... &... &... &... &... &... &...&... \\
\hline                                             
\end{tabular}

\raggedright{\hspace{64pt}$^a$: MR denotes the more metal-rich sample, while MP denotes the more metal-poor sample. See text for details.\\
\hspace{64pt}This table is available in its entirety in machine-readable form. }\\
\end{table*}

\begin{table*}
\caption{N-rich Field Stars Found In This Paper And Paper I.}              
\label{tab:nrich}      
\centering                                      
\begin{tabular}{c c c c c c c c c c }         
\hline\hline  
\# &  RA &  DEC & Teff$^a$ (K) & log(g)$^a$ & [Fe/H]$^a$  & RV$^a$ (km/s) & G\_mag$^b$ (mag) & Distance$^c$ (kpc) & Note\\
\hline 
1  & 260.961223 & 49.579740 & 4530.12 & 1.33 & -1.32 & -206.85 & 11.81 & 5.08  & Paper II\\ 
2  & 272.312586 & 18.683633 & 4588.16 & 1.56 & -1.18 & -109.06 & 12.86 & 5.88  & Paper II\\ 
3  & 126.054337 & 12.348258 & 5111.33 & 2.62 & -1.25 & -137.34 & 14.00 & 3.98  & Paper II\\ 
..  &... &... &... &... &... &... &...&...&... \\
\hline                                             
\end{tabular}

\raggedright{\hspace{34pt}$^a$: From LAMOST DR5 pipeline.\\
\hspace{34pt}$^b$: Gaia DR2 G band magnitudes.\\
\hspace{34pt}$^c$: Distances determined by Chao-Liu.\\
\hspace{34pt}This table is available in its entirety in machine-readable form. }\\
\end{table*}

\begin{figure*}
\centering
\includegraphics [width=0.95\textwidth]{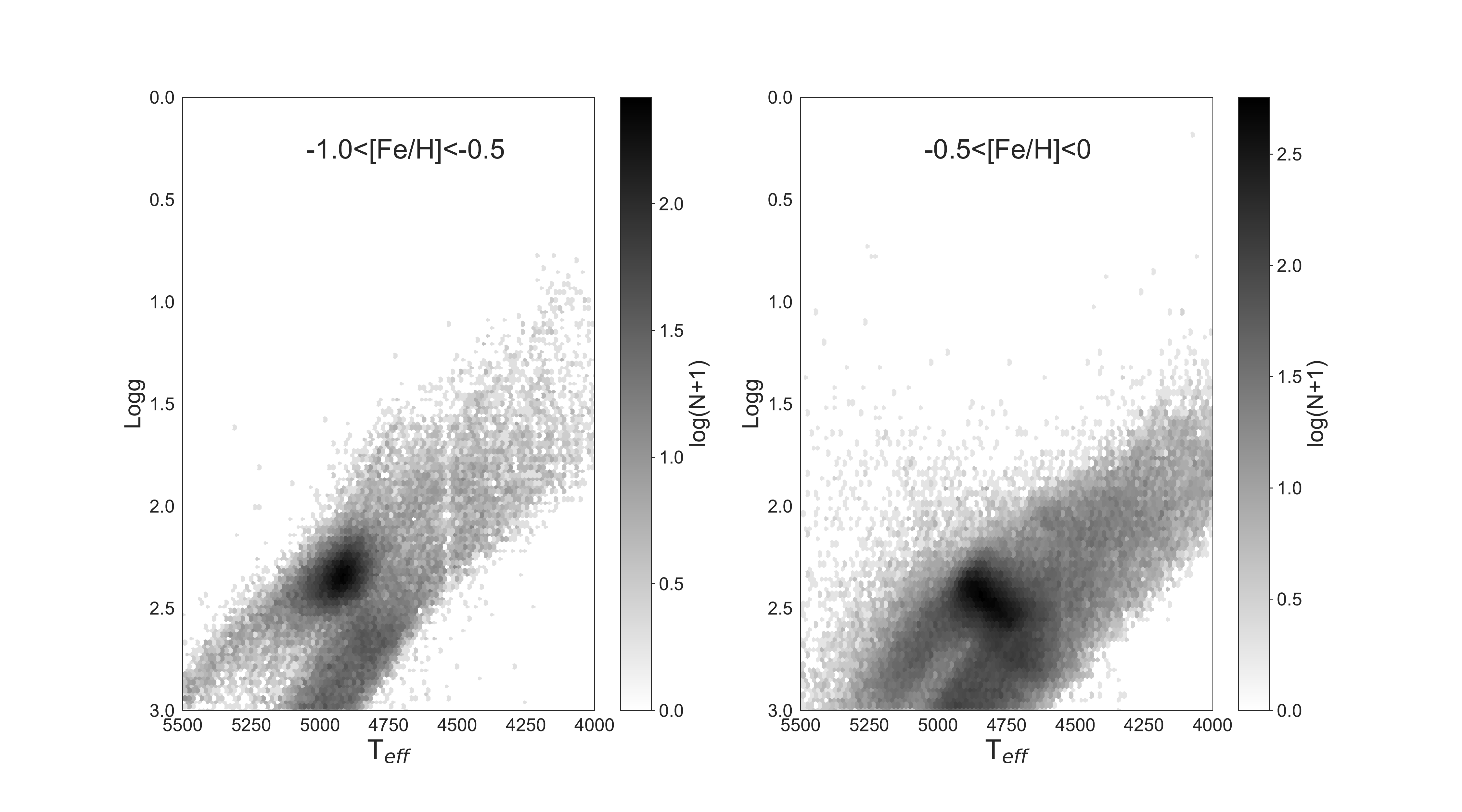} 
\caption{T$_{\rm eff} $ vs. $\log g$ plot for stars with $-1.0<[$Fe/H$]<0$ in LAMOST DR5.
}\label{fig:tloggrc}
\end{figure*}

 In order to study the evolutionary stages of our N-rich field star sample, we plot them along the parent sample in the T$_{\rm eff} $ vs. $\log g$ space (Figure \ref{fig:tlogg}). It is evident that most of our N-rich field stars are located on the RGB, except for a few stars around T$_{\rm eff} \sim 5000-5100$ K and $\log g \sim 2.1-2.2$. What is the evolutionary stage of these stars? We also notice the parent sample shows a feature around T$_{\rm eff} \sim 5250-5400$ K and $\log g \sim 2.5-2.9$, which intersects the RGB. To further explore this, we build two test samples with $-1.0<[$Fe/H$]<-0.5$ (test 1), and $-0.5<[$Fe/H$]<0$ (test 2) from the LAMOST A,F,G,K type star catalog\footnote{T$_{\rm eff} $, $\log g$, and SNR selection criteria are the same as that of our parent sample.}. Figures \ref{fig:tlogg} and \ref{fig:tloggrc} show: (1) The RGB is moving towards the red side as metallicity increases, since high metallicity increases opacity, which causes redder stars; (2) The typical red clump can be found for $-1.0<[$Fe/H$]<0$ stars, while the feature on the low$-$left of the typical red clump is usually referred as the secondary red clump. It is thought to be the results of a smaller He core mass at ignition for stars with $2.2M_{\odot}<M<3.0M_{\odot}$ \citep{Pinsonneault2018}. Therefore, the stars around T$_{\rm eff} \sim 5000-5100$ K and $\log g \sim 2.1-2.2$ are likely red clump stars.

\begin{figure*}
\centering
\includegraphics [width=0.95\textwidth]{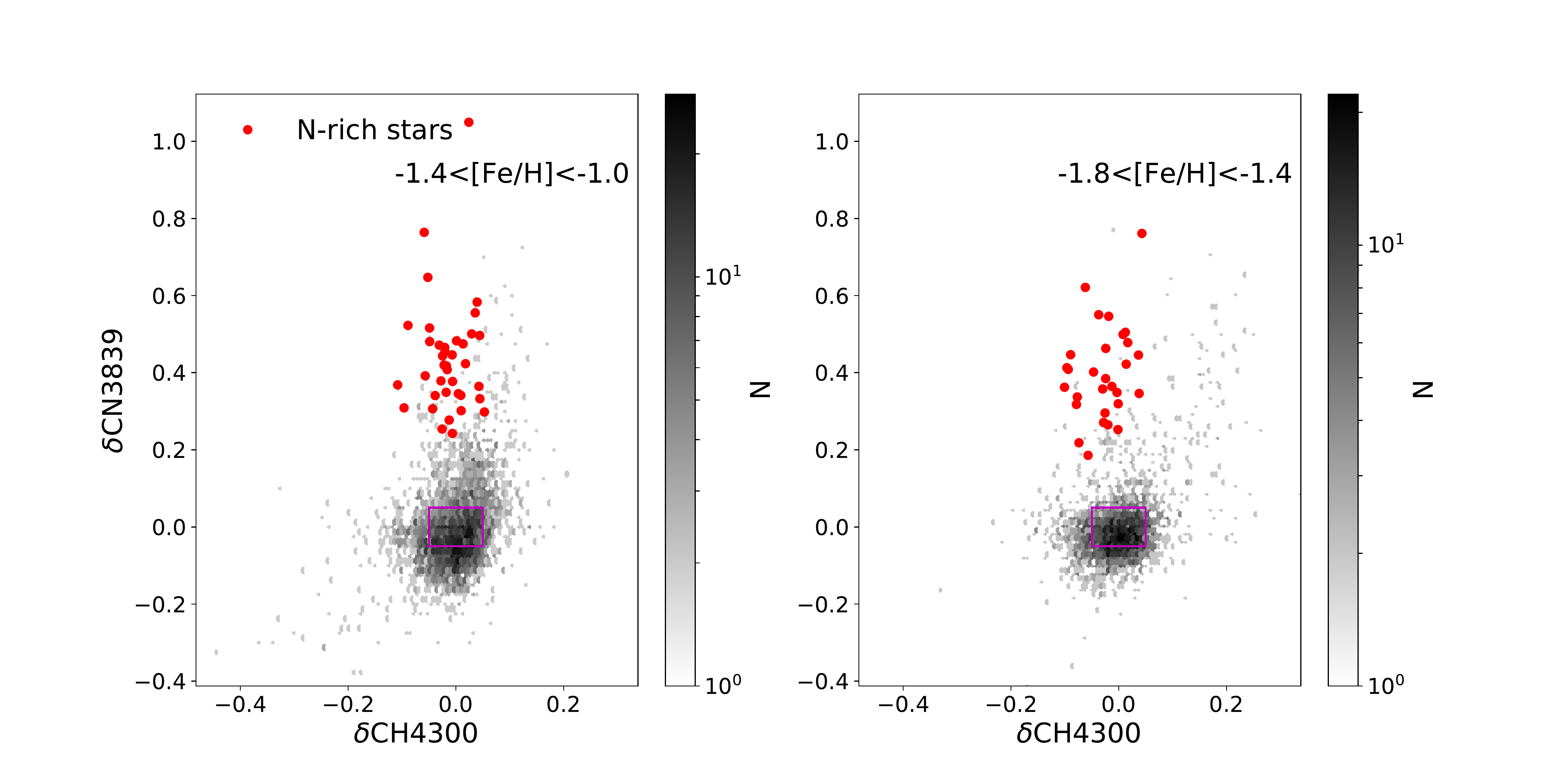} 
\caption{The $\delta$CN3839 vs. $\delta$CH4300 plot. The left panel includes more metal-rich stars, while the right panel includes more metal-poor stars. The N-rich field stars are labelled as red symbols. The gray scale map indicates the number density of metal-poor field stars. A sample of normal metal-poor field stars are selected based on their $\delta$CH4300 and $\delta$CN3839 indices outlined by the magenta box.
}\label{fig:dchdcn}
\end{figure*}

To gain better visualization of the N-rich stars, we define $\delta$CH4300 as the CH4300 index value minus the mean of spectral indices at the T$_{\rm eff}$ of a given star (Table \ref{tab:cnch}). A similar definition is also applied to CN3839. The N-rich stars clearly stand out in the $\delta$CN3839-$\delta$CH4300 plane (Figure \ref{fig:dchdcn}). To define a control sample that represents the normal metal-poor field stars, we select stars that satisfy: (1) $-0.05<\delta$CN3839$<0.05$, (2) $-0.05<\delta$CH4300$<0.05$ (the magenta box in Figure \ref{fig:dchdcn}), and (3) $4000<$T$_{\rm eff}<5000$ K. In total, 1527 stars are selected. We will compare chemical and kinematic properties of N-rich field stars with those of the control sample to discuss their possible origins.

\section{Chemical Abundances}
\label{sect:chem}

\begin{figure}
\centering
\includegraphics [width=0.40\textwidth]{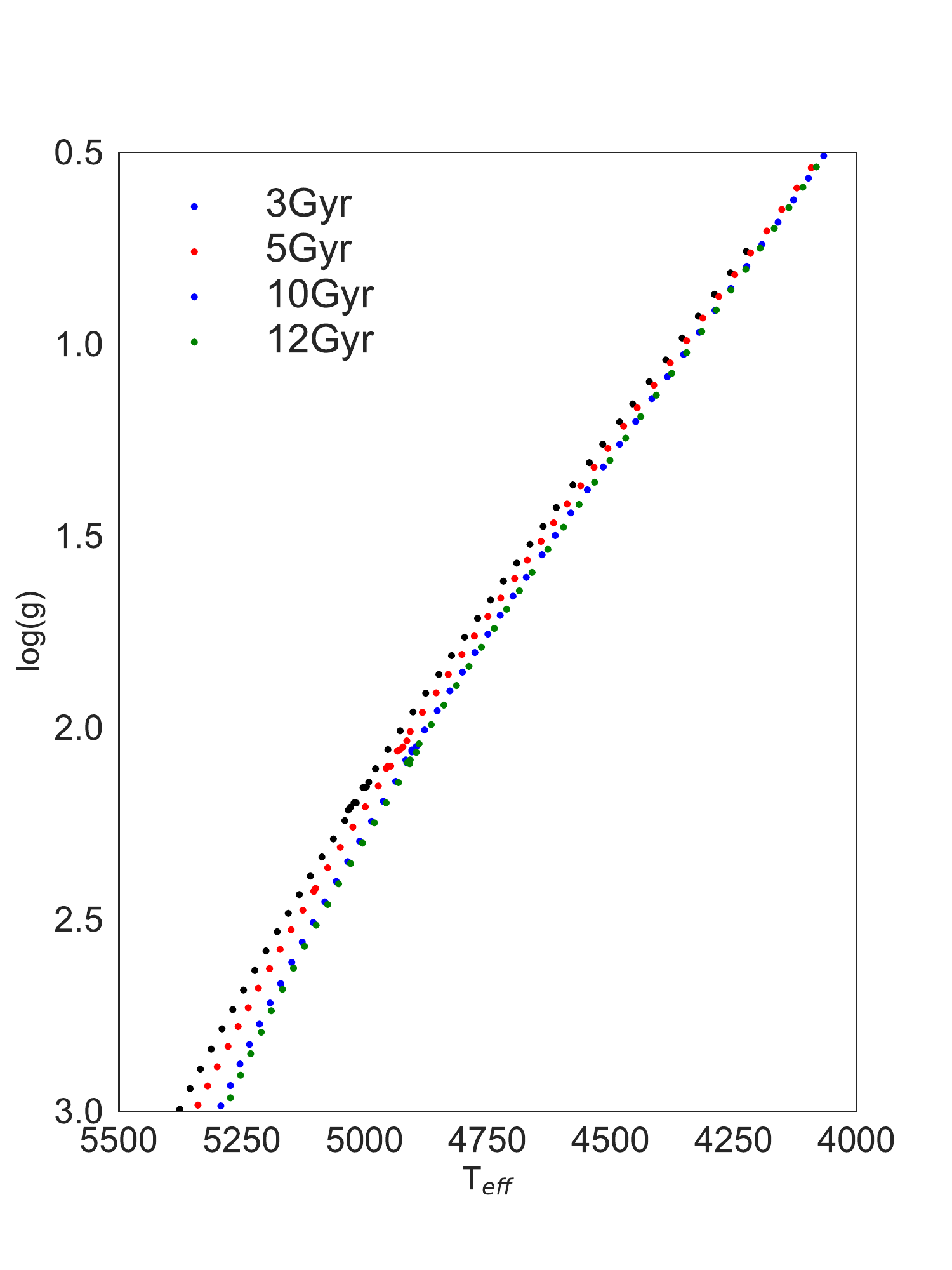} 
\caption{T$_{\rm eff}$ vs. $\log g$ plot of PARSEC isochrones with [Fe/H$]=-1.3$. Isochrone ages are 3 Gyr (black dots), 5 Gyr (red dots), 10 Gyr (blue dots), and 12 Gyr (green dots), respectively.
}\label{fig:par}
\end{figure}

\begin{figure*}
\centering
\includegraphics [width=0.85\textwidth]{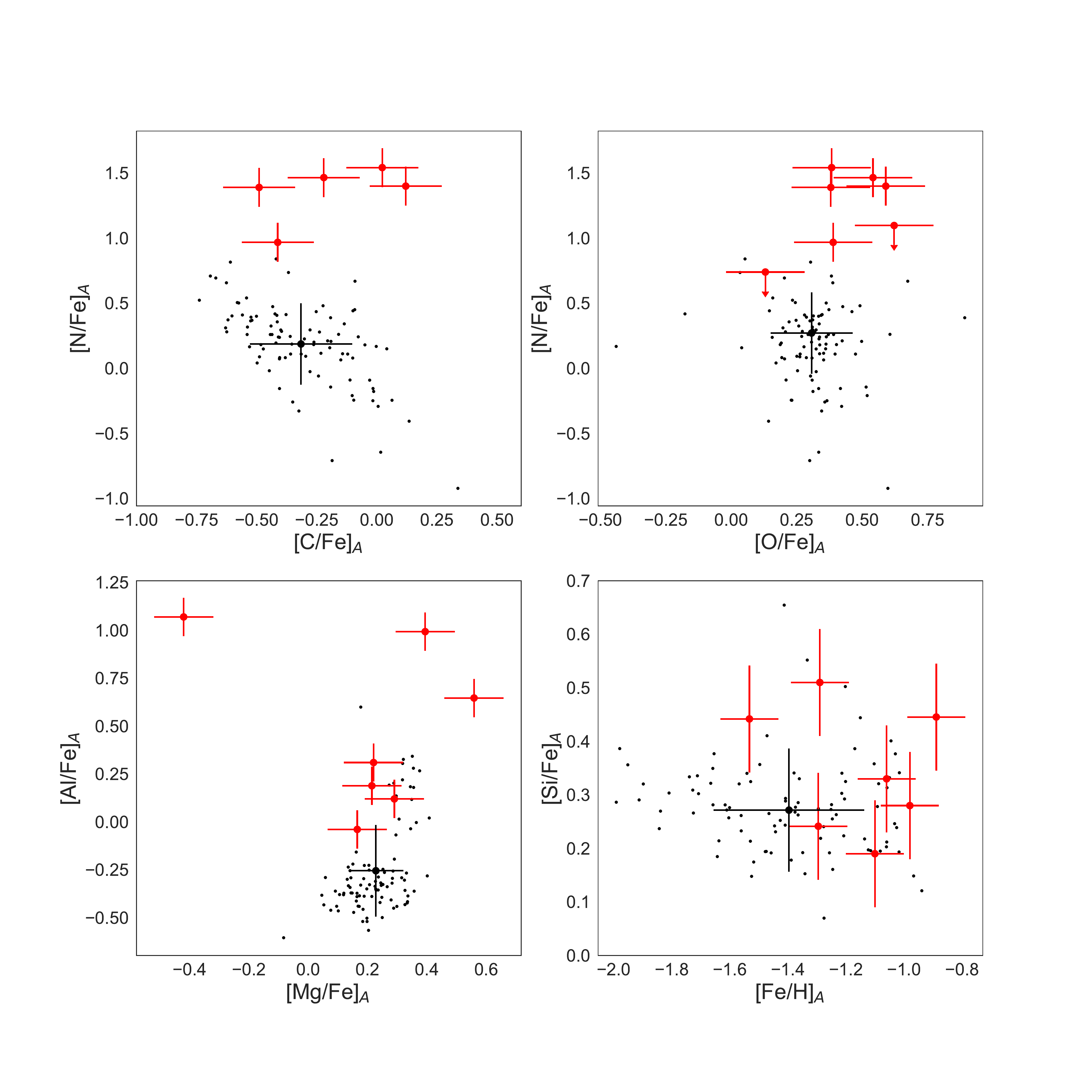} 
\caption{Chemical abundances derived from APOGEE Near-IR spectra. The N-rich field stars commonly observed by LAMOST and APOGEE (DR14) are labelled as red symbols, where the associated errorbars indicate the measurement uncertainties. The chemical abundances of the N-rich field stars are derived using BACCHUS code. The normal metal-poor field stars commonly observed by LAMOST and APOGEE (DR14) are labelled as black dots. Their chemical abundances are given by ASPCAP. The black errorbars indicate their mean and standard deviations. 
}\label{fig:abun}
\end{figure*} 

\begin{figure}
\centering
\includegraphics [width=0.48\textwidth]{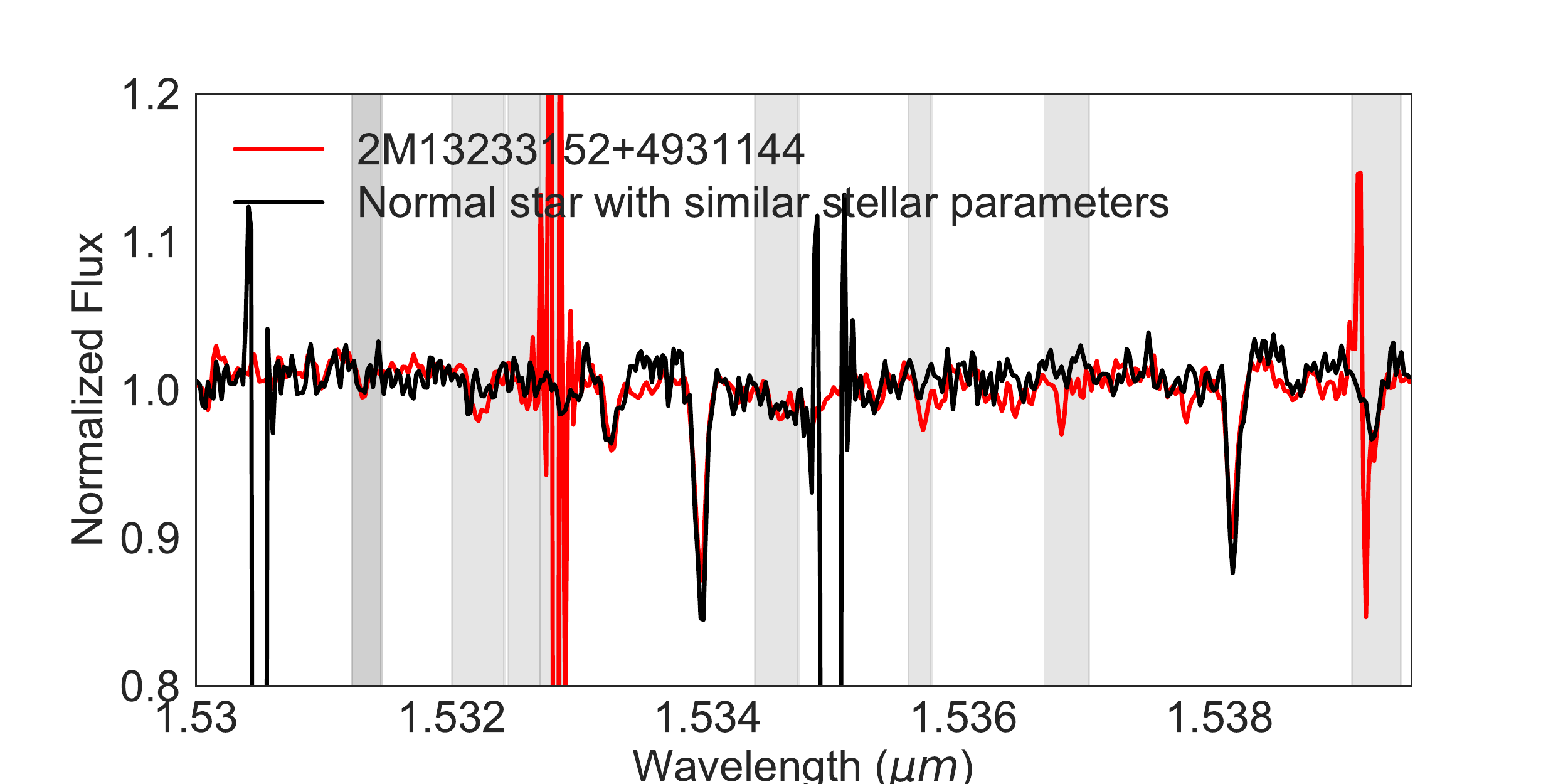} 
\includegraphics [width=0.48\textwidth]{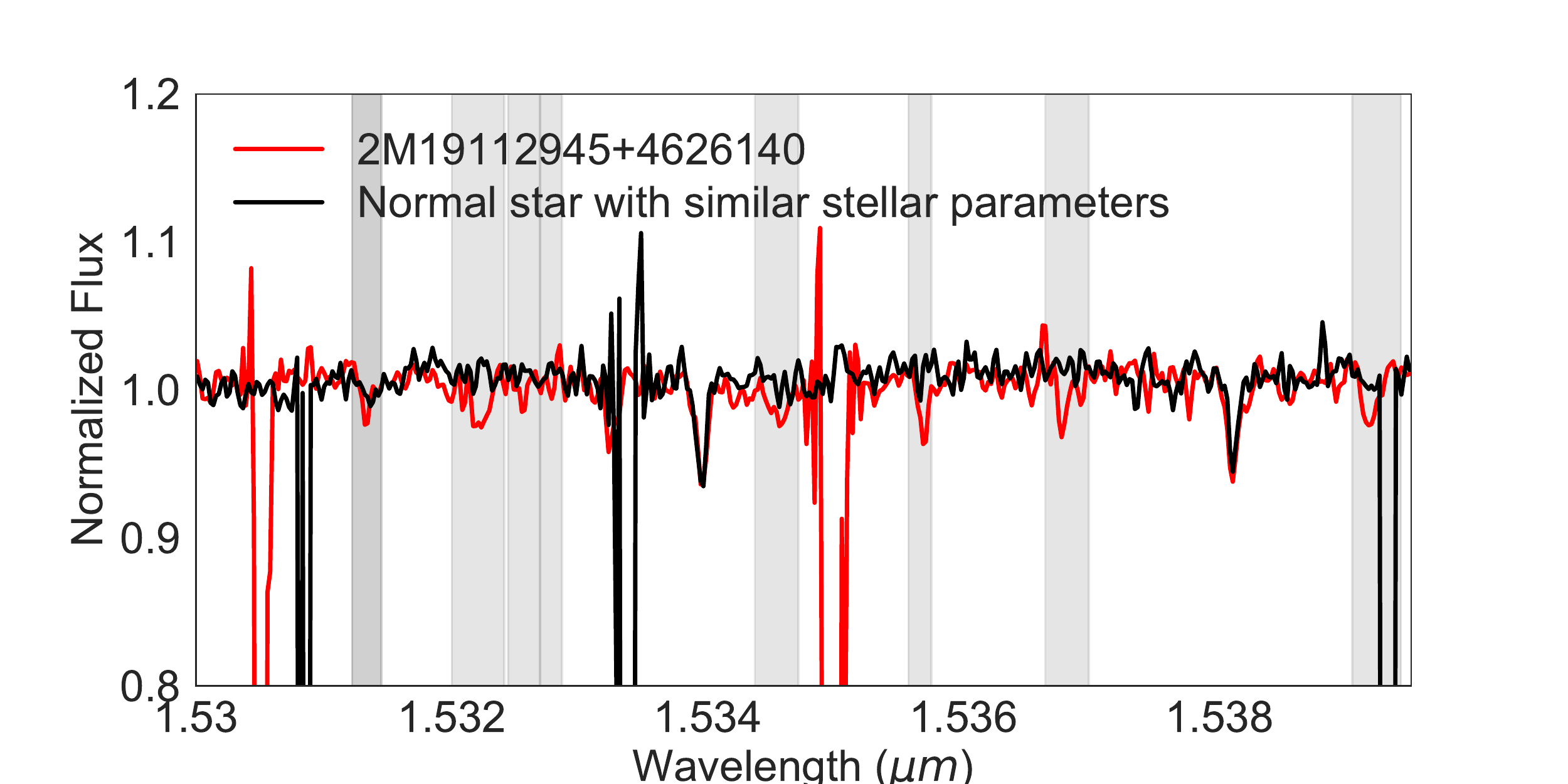} 
\includegraphics [width=0.48\textwidth]{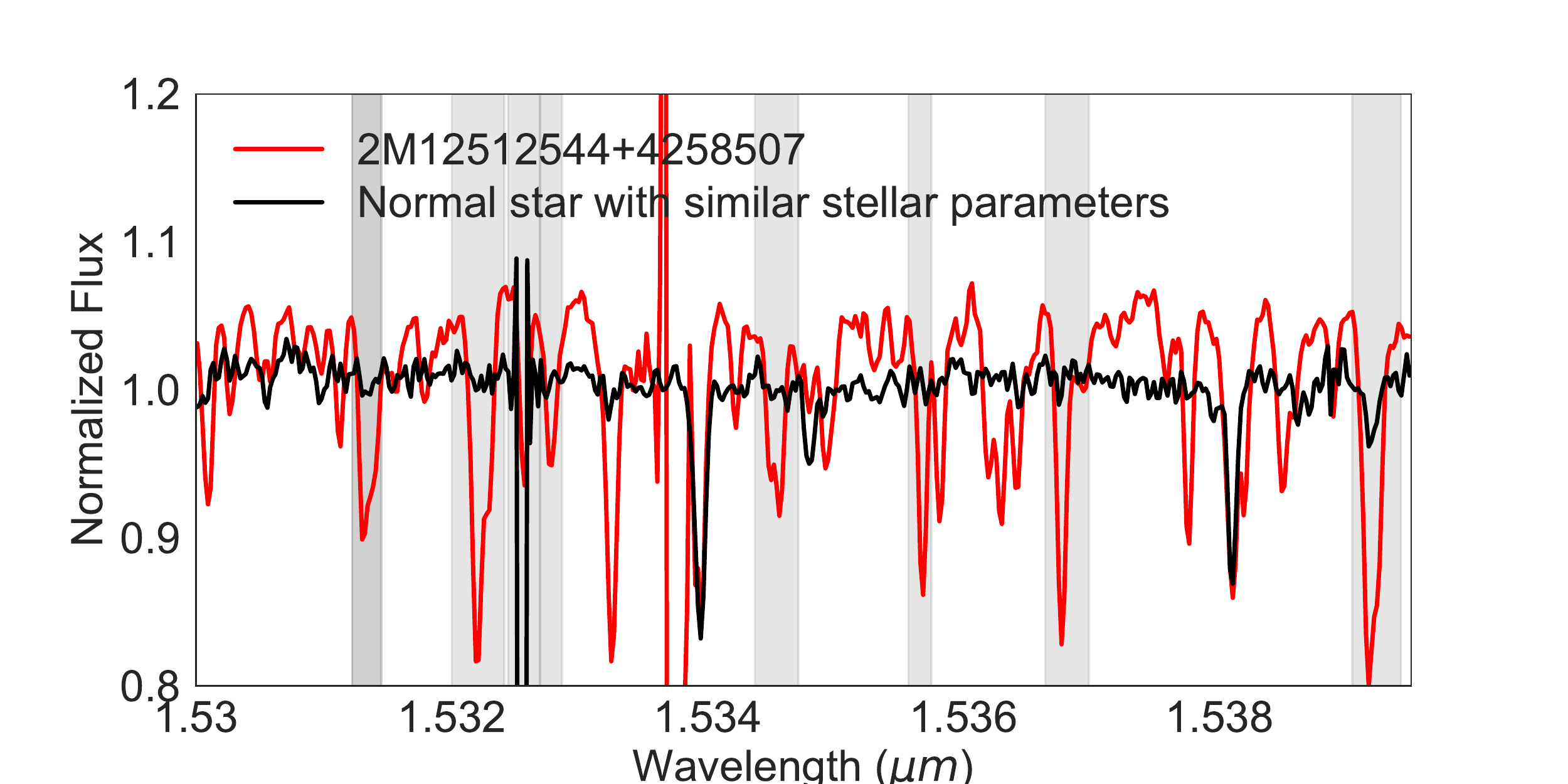} 
\includegraphics [width=0.48\textwidth]{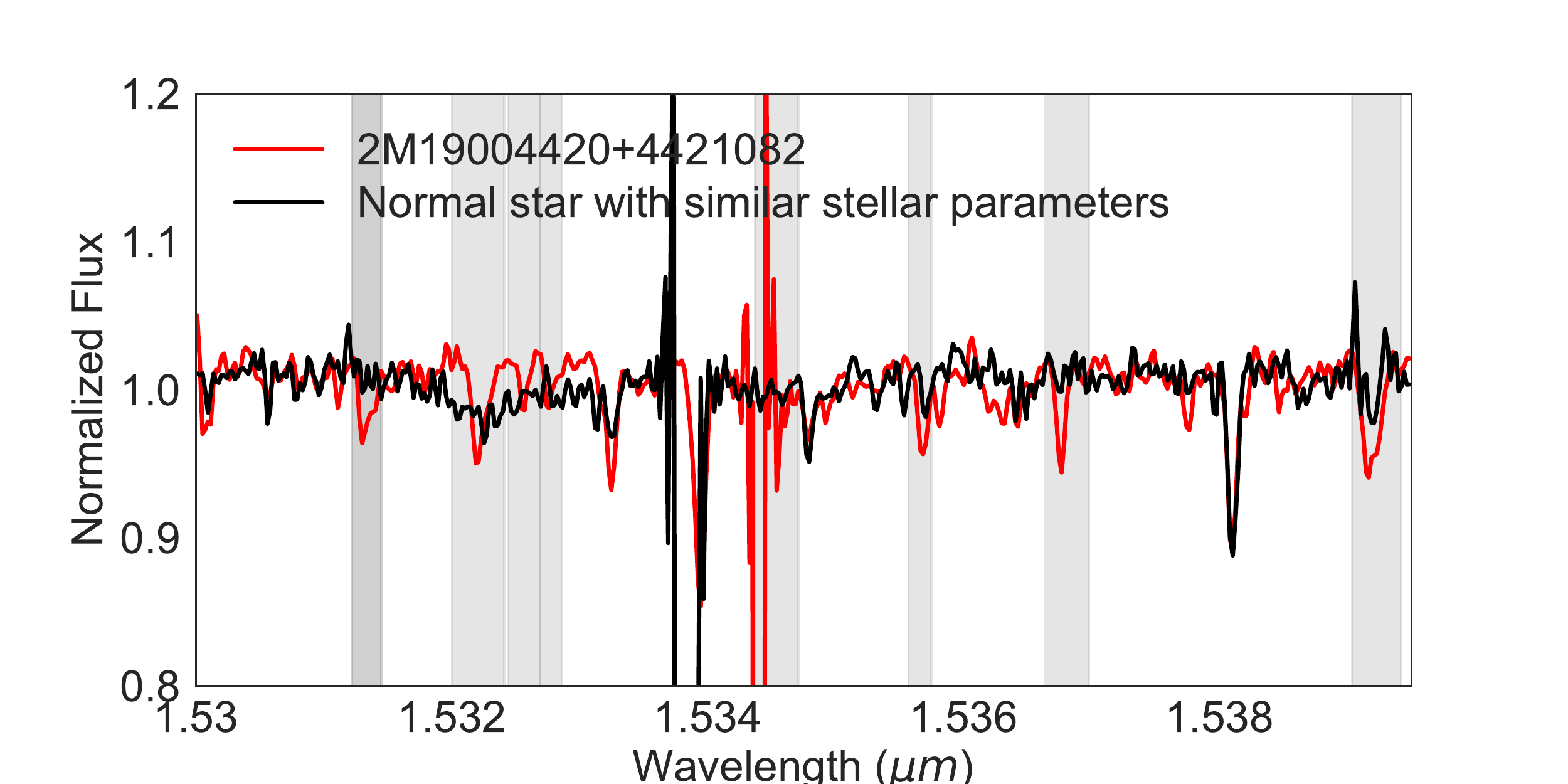} 
\includegraphics [width=0.48\textwidth]{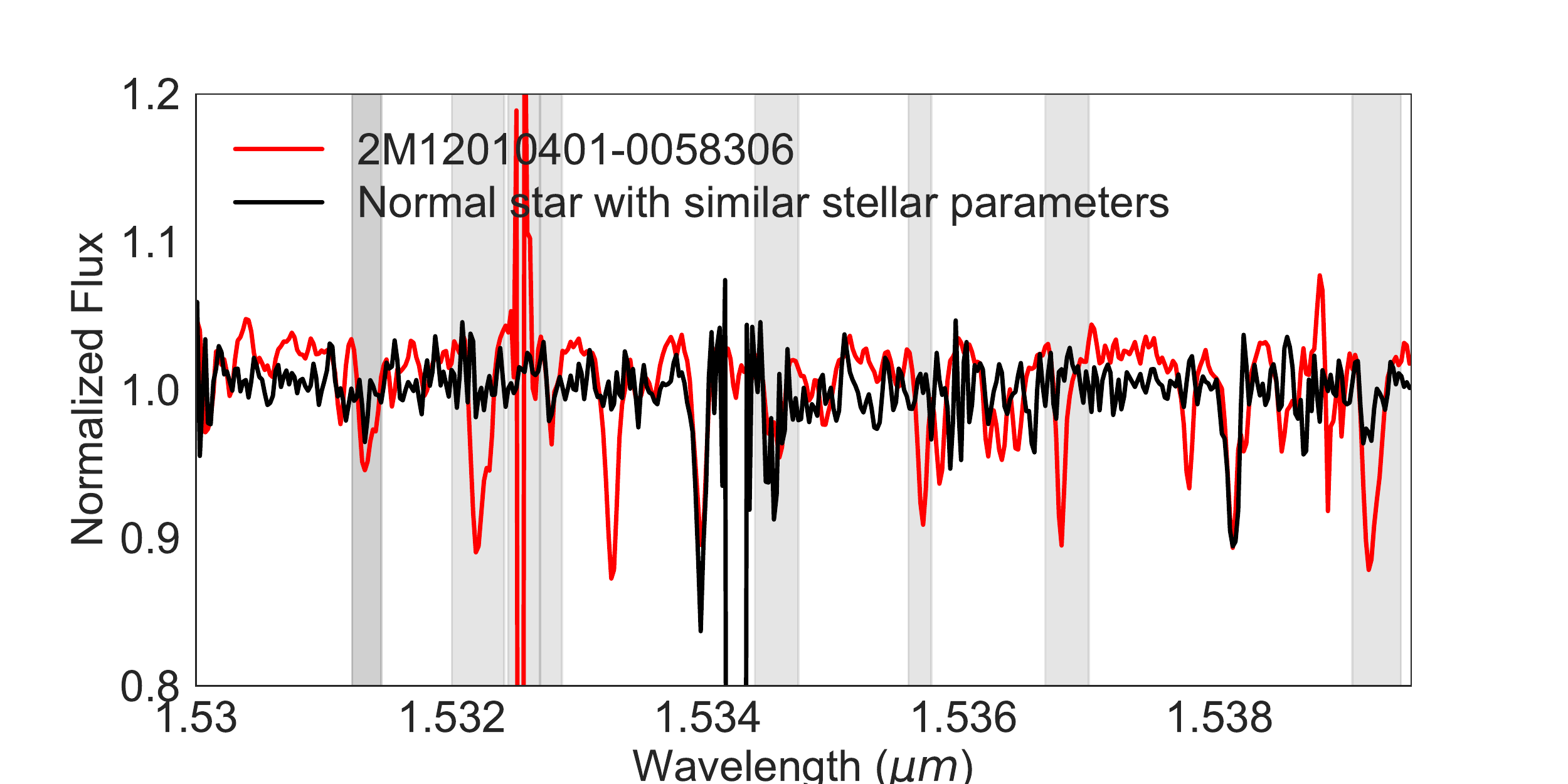} 
\caption{APOGEE spectra covering spectral regions around the $^{12}$C$^{14}$N band (grey vertical bands). The N-rich field stars are labelled as red, while other normal stars with similar stellar parameters are shown as black.
}\label{fig:N1}
\end{figure} 

\begin{table*}
\caption{Common Stars between Our N-rich Field Star Sample and APOGEE DR14.}              
\label{tab:comm}      
\centering                                      
\begin{tabular}{c c c c c c c c c c c c}         
\hline\hline  
APOGEE\_ID & T$_{\rm eff}^a$ (K) & log(g)$^b$ & [Fe/H]$^c$ & [C/Fe]& [N/Fe]& [O/Fe]& [Mg/Fe]& [Al/Fe]& [Si/Fe]& [Ce/Fe]& Note\\
\hline 
2M12561260+2804017  & 4848.65 & 1.87 & -1.30 & ... & 1.10 & 0.62 & 0.17 & -0.04 & 0.24 & 0.59  & Paper I\\ 

2M13233152+4931144  & 4778.95 & 1.89 & -1.10 & -0.41 & 0.97 & 0.39 & 0.21 & 0.19 & 0.19 & 0.09  & Paper I\\ 

2M19112945+4626140  & 4867.39 & 1.80 & -1.53 & -0.22 & 1.47 & 0.54 & 0.39 & 0.99 & 0.44 & 0.41  & Paper II\\ 

2M12512544+4258507  & 4753.45 & 2.26 & -0.89 & 0.02 & 1.54 & 0.38 & 0.56 & 0.65 & 0.45 & 0.92  & Paper II\\ 

2M19004420+4421082  & 4823.72 & 2.05 & -0.98 & -0.49 & 1.39 & 0.38 & 0.22 & 0.31 & 0.28 & 0.56  & FT19\\ 

2M12010401-0058306  & 4880.99 & 2.25 & -1.06 & 0.12 & 1.40 & 0.59 & 0.29 & 0.12 & 0.33 & 0.92  & FT19\\ 

2M15535831+4333280  & 4656.55 & 1.58 & -1.29 & ... & 0.74 & 0.13 & -0.42 & 1.07 & 0.51 & 0.32  & FT19\\ 

\hline                                             
\end{tabular}

\raggedright{\hspace{0pt} $^a$: Photometric T$_{\rm eff}$; $^b$: Determined from 10 Gyr isochrones; $^c$: ASPCAP values. Other  abundances are derived using BACCHUS.}\\
\end{table*}

The idea of ``chemical tagging'' \citep{Freeman2002} envisages chemical abundances as labels for a given star which tell its evolution history, and may be used as tracers to help us constrain its birth environment.  In this sense, chemical abundances are crucial to understand the N-rich star origins. 

In spite of observationally more expensive, high-resolution spectra give more accurate chemical abundances. Thus, we seek help from high spectral resolution surveys, like APOGEE \citep{Majewski2017}.
After cross-matching with the APOGEE DR14 database, we find 97 common stars in the control sample and 5 common stars in our N-rich star sample. APOGEE provides up to more than 20 elemental abundances \citep{Holtzman2015} with the APOGEE Stellar Parameter and Chemical Abundances Pipeline (ASPCAP; \citealt{GP2016}). While ASPCAP results are statistically reliable to $~0.1$ dex for several elements \citep{Holtzman2015}, cautions should be taken when using ASPCAP results for individual stars \citep[e.g.,][]{SU2018}, especially for metal-poor stars (below [Fe/H$]=-1.4$ dex). Given that accurate chemical abundance derivation is computationally expensive, we plan to (1) use ASPCAP results of these 97 common stars in the control sample without examining the details of each spectrum, but we mainly discuss their statistical mean and standard deviation values. (2) derive detailed chemical abundances for common stars in the N-rich field star sample. It turns out that
 3 of the 5 N-rich common stars are listed in \citet[][FT19]{FT2019c}. In this paper, we use the chemical abundances given by FT19, and apply the same procedure of deriving chemical abundances with photometric T$_{\rm eff}$ to the other two APOGEE-LAMOST common stars. We also include the two N-rich field stars with APOGEE spectra from Paper I. 
 
Why are there four N-rich field stars observed by APOGEE but not included in FT19? It turns out that they are either outside the metallicity range selected in FT19, or right on the border that separates N-rich and normal field stars in FT19. Furthermore, we also check for common stars in \citet{Carretta2010, Martell2010, Martell2011, Ramirez2012, Lind2015, Martell2016, Fernandez-Trincado2016, Fernandez-Trincado2017a, Schiavon2017chem}. No overlap is found, therefore, we are presenting 53 newly identified N-rich stars in this work. 
 
Briefly speaking,  we first calculate the photometric T$_{\rm eff}$ with $J_{2MASS}-K_{s, 2MASS}$ color using the correlation of \citet{GH2009}, where the color is extinction-corrected using the E(B-V) provided by the APOGEE team. Photometric $\log g$ are estimated from 10 Gyr PARSEC isochrones with ASPCAP [Fe/H] (Table \ref{tab:comm}). Then we estimate the chemical abundances using the Brussels Automatic Stellar Parameter (BACCHUS) code \citep{Masseron2016}. Readers are referred to \citet{Tang2018} and FT19 for detailed description about how we setup the BACCHUS code to compute chemical abundances. 
The results from the BACCHUS code are listed in Table \ref{tab:comm}. It is worth noticing that when [C/Fe] is too weak to determine a reasonable value (two out of the seven common stars), the associated [N/Fe] is in fact an upper limit. This is related to how we determine the C, N, and O abundances: we first derive $^{16}$O abundances from $^{16}$OH lines, then derive $^{12}$C from $^{12}$C$^{16}$O lines and $^{14}$N from $^{12}$C$^{14}$N lines. If $^{12}$C$^{16}$O lines are too weak, then we fix [C/Fe] to solar value, so the associated [N/Fe] should be interpreted as upper limit, and used with caution.
 There is concern about adopting 10 Gyr isochrones to evaluate our $\log g$, which may propagate errors to our abundance measurements. We download 3, 5, 10, and 12 Gyrs isochrones with a typical metallicity of our sample: [Fe/H$]=-1.3$ from the PARSEC isochrone website\footnote{http://stev.oapd.inaf.it/cgi-bin/cmd}. The $\log g$ of different age isochrones in fact vary less than 0.1 dex at a given T$_{\rm eff}$ (Figure \ref{fig:par}). This uncertainty in $\log g$ is much less than the $\log g$ uncertainty that we assume in FT19 and this paper (0.36 dex). Therefore, the abundance uncertainties caused by adopting different age isochrones are negligible compared to the uncertainties presented in Table 3 of FT19.

The BACCHUS-derived chemical abundances for our N-rich field stars and the ASPCAP-derived chemical abundances for the control sample are shown in Figure \ref{fig:abun}. On the upper-left panel, as we have demonstrated in Paper I, we find normal metal-poor field stars are going through the so-called extra-mixing \citep{Iben1967, Gratton2000}, and our N-rich field stars are clearly above this sequence. Therefore, the N-enrichment cannot be explained by the classical extra-mixing theory. A revision of the extra-mixing theory, or new nucleo-synthetic process that can produce very high N abundances is needed.  Two issues concerning the data reduction may complicate our discussion here: first, the N abundances derived by ASPCAP for APOGEE DR14 do not reach [N/Fe$]>1.0$, due to the limited grid of models \citep{Masseron2019}; second, $^{12}$C$^{14}$N spectral features in the APOGEE spectra become weak for stars with high temperature. For example, \citet{Masseron2019} suggested that [N/Fe] for stars with T$_{\rm eff}$ above 4600 K are mostly upper limits. To prove that our APOGEE common N-rich field stars are truly N-enhanced, we show the APOGEE spectra of our N-rich field stars (without [N/Fe] upper limits) in Figure \ref{fig:N1}. We center the spectra on the regions around the $^{12}$C$^{14}$N band to visualize the spectral features. Clearly, all the N-rich field stars show stronger spectral absorption of the $^{12}$C$^{14}$N band, compared to other normal stars with similar stellar parameters. Our LAMOST-APOGEE common stars are bona fide N-rich field stars. This demonstrates our ability to select bona fide N-rich field stars from low resolution LAMOST spectra, and the purity of our N-rich field star sample. 

On the upper-right panel, the N-rich field stars show similar, or even slightly higher, O abundances as other normal metal-poor field stars. Classical extra-mixing is believed to affect C and N yields but not the C$+$N$+$O yields \citep{Masseron2019}. The C$+$N$+$O abundance sum for our five N-rich field stars with complete C, N, O abundances  is 8.44$\pm$0.23, while for other normal metal-poor field stars, it is 7.64$\pm$0.29. To minimize the effect of metallicity, we also compute the [C$+$N$+$O/Fe] for both samples. It is 0.68$\pm$0.11 for N-rich field stars and 0.20$\pm$0.13 for normal metal-poor field stars. Therefore, our N-rich field stars clearly have higher C$+$N$+$O yields compared to normal metal-poor field stars. This supports our above statement that the classical extra-mixing theory cannot explain the N-enhancement that we found. As discussed in several recent literature \citep{Villanova2010, Yong2015, Masseron2019}, whether GC stars of different generations show distinct C$+$N$+$O yields is still unclear; even for the same GC (NGC 1851), whether the two generations show different C$+$N$+$O abundance sum is still inconclusive \citep{Villanova2010, Yong2015}. Thus, directly linking the different C$+$N$+$O abundance sum of our N-rich field stars compared to normal metal-poor field stars with the GC escapee scenario is still premature.

On the lower-left panel, we find that [Al/Fe] of the N-rich field stars are clearly higher than that of the control sample. Some can even reach [Al/Fe$]\sim 1$. [Mg/Fe] are generally enhanced to a level of about 0.3 dex for both samples, consistent with their metal-poor halo star nature. Exception is found in 2M15535831$+$4333280, where [Mg/Fe] is as low as $-0.42$ dex. This star has been presented in FT19, and similar Mg-depleted N-rich field stars were also discussed in \citet{Fernandez-Trincado2016, Fernandez-Trincado2017a}. N-rich field stars showing Mg depletion are in fact rare, but whether Mg-depleted and Mg-enhanced N-rich field stars have common origins is still unclear. On the lower-right panel, [Si/Fe] of the N-rich field stars are scattered around $0.2-0.5$ dex, generally consistent with other normal metal-poor field stars. Finally, we notice that the s-process element ([Ce/Fe]) abundances of two N-rich field stars are large ($\sim 0.9$ dex),  pointing to a possible AGB enriched scenario (Section \ref{sect:AGB}). 
We are aware of the limited sample size of our N-rich stars with high-resolution chemical abundances, and a campaign to obtain high-resolution spectra for 20-30 N-rich field stars is on the way.

\section{Tracking Orbits}
\label{sect:orb}

\begin{figure}
\centering
\includegraphics [width=0.45\textwidth]{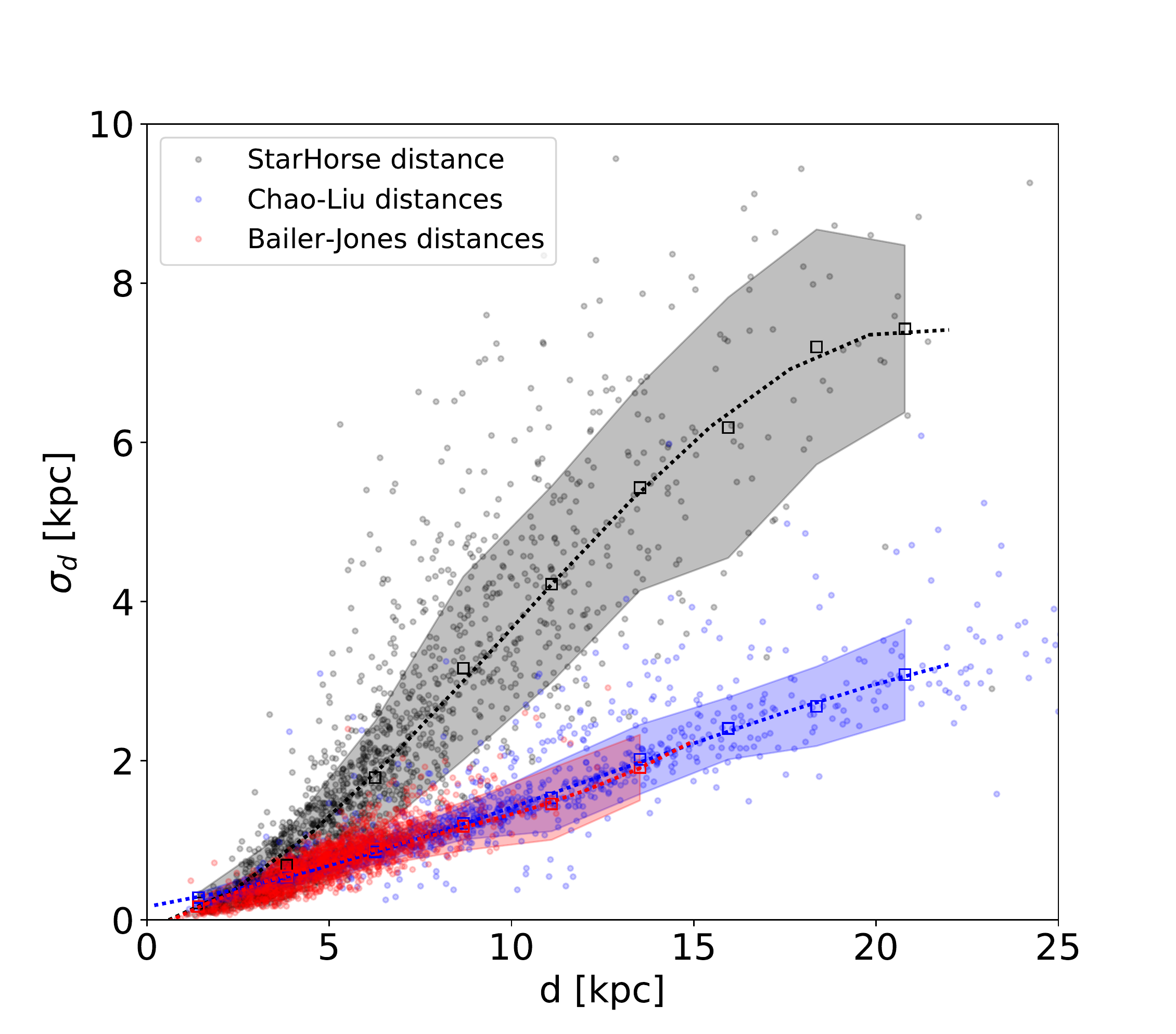} 
\caption{Distance error as a function of distance value given by three different methods for all N-rich field stars and normal metal-poor field stars. LC distances are labelled as blue dots, SH distances are grey dots, BJ distances are red dots. The mean errors at given distances and the associated 1 $\sigma$ uncertainty regions are shown as dashed lines and shaded regions. (see text). 
}\label{fig:distcomp}
\end{figure} 

\begin{figure*}
\centering
\includegraphics [width=0.95\textwidth]{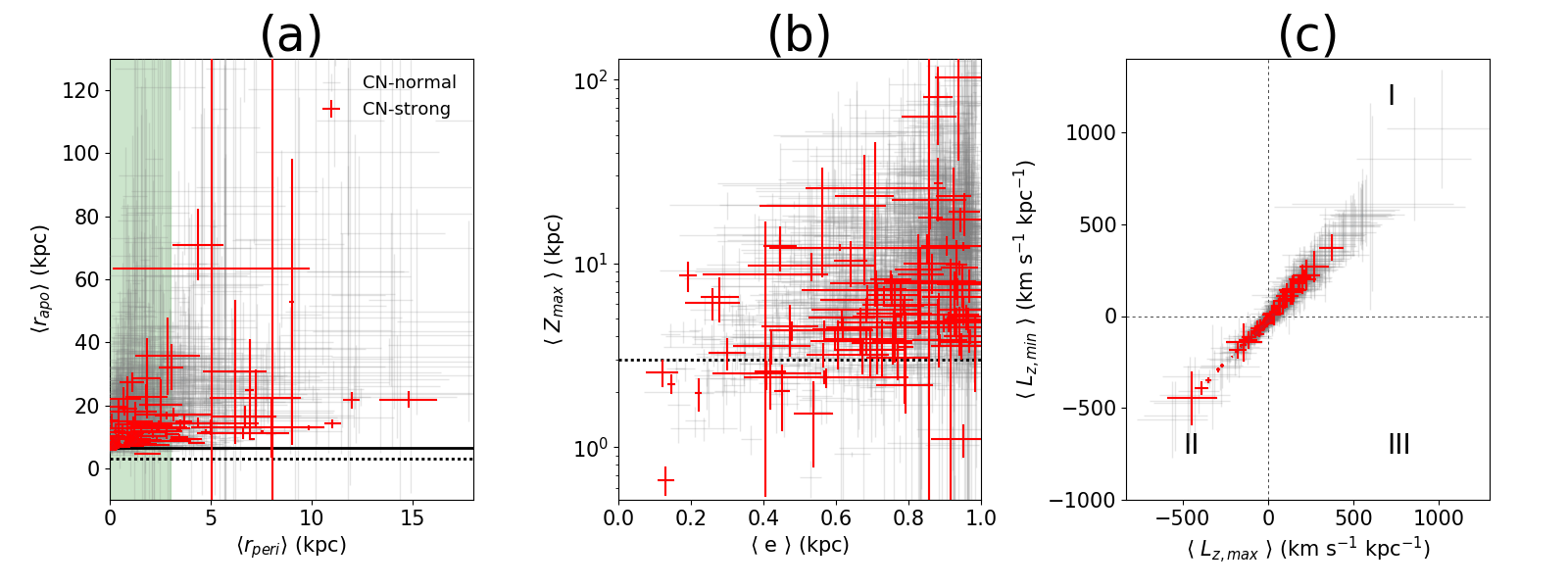} 
\caption{The plots of average values of orbital parameters over 1000 realizations for the N-rich field stars (red symbols) and the control sample (gray symbols). The parameters include: perigalactic radius ($<r_{\rm peri}>$), the apogalactic radius ($<r_{\rm apo}>$), the orbital eccentricity ($<e>$), the maximum vertical amplitude ($<Z_{\rm max }>$), the minimum angular momentum in z direction ($<L_{z,min}>$), and the maximum angular momentum in z direction ($<L_{z,max}>$). The error bars indicate uncertainty ranges given by the 16th and 84th percentile values.
}\label{fig:paraall}
\end{figure*} 

\begin{figure*}
\centering
\includegraphics [width=0.48\textwidth]{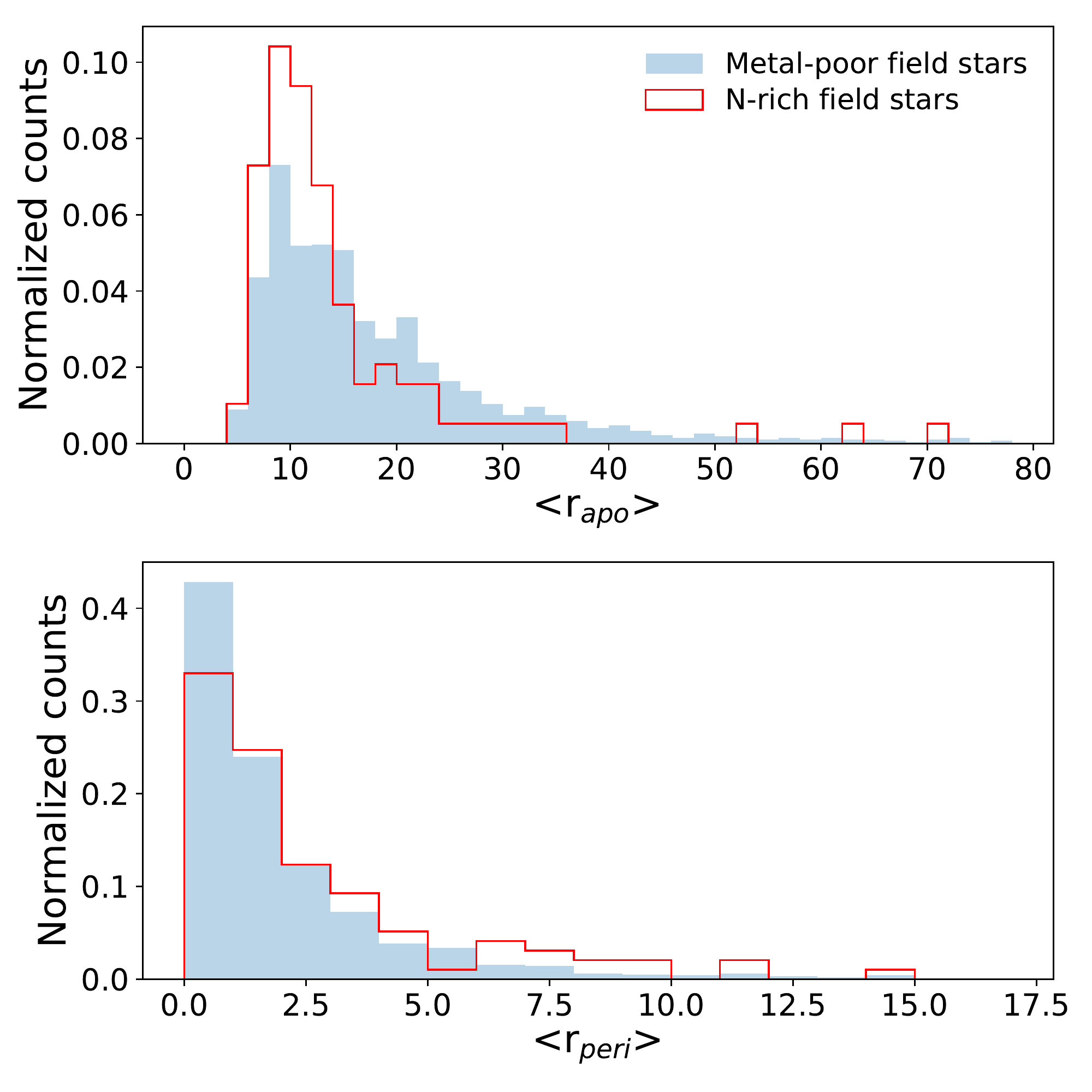} 
\includegraphics [width=0.48\textwidth]{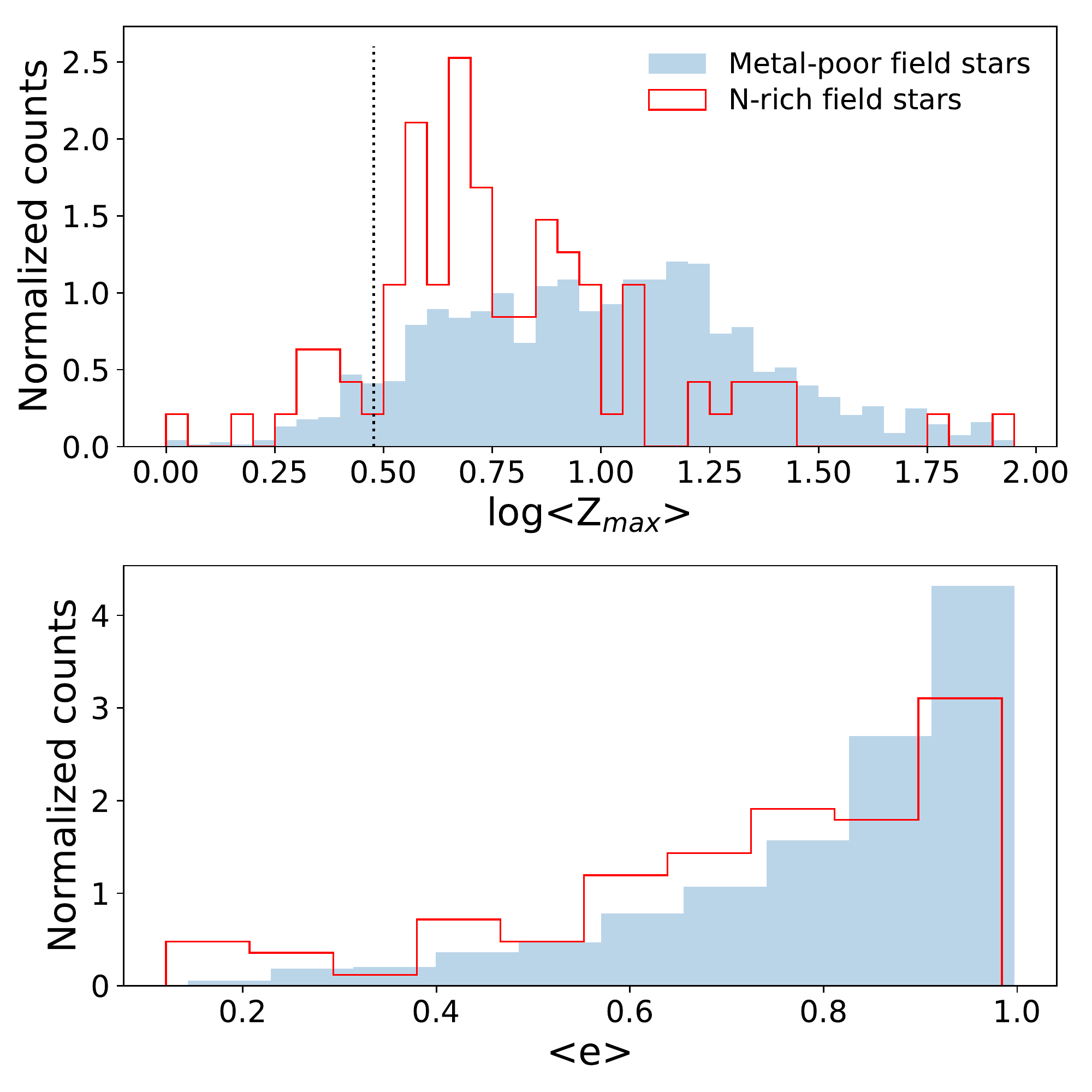} 
\caption{The $<r_{\rm apo}>$ (upper-left panel), $<r_{\rm peri}>$ (lower-left panel), $\log <Z_{\rm max }>$ (upper-right panel) and $<e>$ (lower-right panel) distribution histograms of N-rich field stars (red histograms) and normal metal-poor field stars (gray histograms). 
}\label{fig:zmaxe}
\end{figure*} 

As stars travel through the MW gravitational potential, their orbital energy and angular momentum are generally conserved for a few Gyrs, given the slowly varying MW gravitational potential. The orbital parameters of our N-rich field stars, especially when compared with normal metal-poor field stars, may reveal their unknown past. 
To provide a comprehensive dynamical study of our sample, we simulate the Galactic orbits in a 3-D steady-state gravitational potential model for the Galaxy, modeled as the sum of axisymmetric and non-axisymmetric components. The axisymmetric component is made up of the superposition of many composite stellar populations belonging to the thin disk (7 stellar sub-populations with ages ranging between 0.15 to 10 Gyr), where the density profile of each component follow the Einasto laws \citep{Einasto1979} and observationally constrained as presented in \citet{Robin2003}. The model also considers the contribution by two thick disks (with ages of 10 and 11 Gyr) whose profiles follow the sech$^{2}$ laws similar to that presented in \citet{Robin2014}, and an interstellar matter (ISM) component with the density profile as presented in \citet{Robin2003}. The model, also correctly accounts for the underlying stellar halo, modeled by a Hernquist profile as already described in \citet{Robin2014}. The non-axisymmetric component is modeled by a ``boxy/peanut'' bar structure whose density profile is observationally constrained from 2MASS data (see \citealt{Robin2012}), with an assumed mass, present-day orientation and pattern speeds, which are within observational estimates: Bar mass $-$ 1.0$\times$10$^{10}$ M$_{\odot}$ \citep{Portail_2017}; Angle $-$ 20$^{\circ}$ (present-day angular offset from the Galactic x-axis in the direction of rotation); and $\Omega_{\rm bar}-$ 33 to 53 km s$^{-1}$ kpc (in increments of 10 km s$^{-1}$ kpc, which has a co-rotation radius of $\sim$ 5.5 $-$ 6.5 kpc) in line with \cite{Fernandez-Trincado2017Thesis}, respectively. All these stellar components are surrounded by an isothermal dark matter halo component with a density mass as presented in \citet{Robin2003}. It is important to note that the literature listed here indicates the origin of the density profiles of our model, but the mathematical functions associated to the gravitational potential ($\Phi(x,y,z)$) will be presented for the first time in a forthcoming paper by Fern\'andez-Trincado et al. (in preparation). Efforts are underway to provide our code to the community by running it on a public server (\url{https://gravpot.utinam.cnrs.fr}).

To the Galactic orbits, we adopt a solar position of R$_{\odot}$ = 8.3 kpc, Z$_{\odot}$ = 11 pc, local kinematic parameters of V$_{LSR}$ $=$ 239 km/s (for the motion of the local standard of rest) and [U,V,W]$_{\odot}$ = $[-11.10, -12.24, 7.25]$ km s$^{-1}$, in line with \citet{Brunthaler2011}. We use a right-handed, Cartesian Galactocentric coordinate system, where the X-axis is oriented toward  $ l =$ 180$^{\circ}$, the Y-axis is oriented toward $l =$ 270$^{\circ}$, and the disk rotates toward $l \sim 90^{\circ}$. For the computation of Galactic orbits, we have employed a simple Monte Carlo approach and the Runge-Kutta algorithm of seventh-eight order. The uncertainties in the input data (e.g., $\alpha$, $\delta$, distance, proper motions, and line-of-sight velocity errors) were randomly propagated as 1$\sigma$ variation in a Gaussian Monte Carlo re-sampling. For each star, we computed a thousand orbits, computed backward in time during 3 Gyr. The average value of the orbital elements was found for our 1000 realizations, with uncertainty ranges given by the 16th and 84th percentile values.  

The input data for the Galactic model: distances, RVs, and absolute proper motions, are retrieved from the most recent survey results. We use absolute proper motions from the latest Gaia DR2 \citep{Brown2018, Katz2018}, and RVs from LAMOST. The typical uncertainty of LAMOST RVs is 4 km/s, while the typical uncertainty of absolute proper motions is 0.05 mas/yr. We compare distances derived from three different methods: (1) Bayesian spectro-photometric distances with no assumptions about the underlying populations \citep{Carlin2015}, hereafter LC distances (led by Chao Liu); (2) Bayesian spectrophotometric distances with flexible Galactic stellar-population priors \citep{Queiroz2018}, hereafter SH (StarHorse) distances; (3) Bayesian Gaia DR2 parallax-based distances \citep{BailerJones2018}, hereafter BJ distances. We compare the distance errors of the aforementioned methods in Figure \ref{fig:distcomp}. We notice that (1) uncertainties of SH distances are the largest among the three; (2) BJ distances and LC distances have comparable uncertainties for nearby stars (distance less than 7 kpc), but the number of stars with distance greater than 7 kpc is almost negligible if BJ distances are assumed. Because BJ distances are based on parallaxes from Gaia DR2, if a star is too far away, the parallax becomes too small to be detectable. Currently, the parallax-based distances of stars further than $\sim 5$ kpc are expected to be dominated by their assumed priors. Therefore, we adopt LC distances to compute stellar orbits in this work.

From the integrated set of orbits, we compute (1) the perigalactic radius, $r_{\rm peri}$, (2) the apogalactic radius, $r_{\rm apo}$, (3) the orbital eccentricity, defined as $e=(r_{\rm apo} - r_{\rm peri})/(r_{\rm apo} + r_{\rm peri})$, (4) the maximum vertical amplitude, $Z_{\rm max }$.

A long list of studies in the literature has presented different ranges for the bar pattern speeds \citep{Portail_2017, Monari_2017a, Monari_2017b}. For our computations, 
we assume three pattern speeds, $\Omega_{\rm B}=33, 43, 53$ km s$^{-1}$ kpc$^{-1}$. Since most of our stars are located in the halo, we do not expect substantial differences in our orbital parameters when adopting different pattern speeds, as we found in Paper I. Therefore, we adopt $\Omega_{\rm B}=43$ km s$^{-1}$ kpc$^{-1}$ in the following study. 

In this work, we have combined the N-rich field stars in this paper and Paper I to form a $\sim100$ star sample. This is the largest homogeneous N-rich field star samples with kinematic and orbital information. Readers are referred to Paper I for the classical Toomre diagram. In this work, we use histograms to better visualize the distribution differences between samples. Figures \ref{fig:paraall} and \ref{fig:zmaxe} show that our N-rich field stars have smaller $<r_{\rm apo}>$, $<e>$, and $<Z_{\rm max }>$ compared to the control sample.  (1) The N-rich field star sample and the control sample have no star with $<r_{\rm apo}>$ less than 3 kpc, indicating no star is constrained inside the bulge \citep{Barbuy2018}. (2) Two samples have more stars towards high eccentricity. (3) The two sample Kolmogorov-Smirnov (K-S) test indicates that the probability that  $<Z_{\rm max }>$ (and $<r_{\rm apo}>$) of the two samples are drawn from the same parent population is lower that $10^{-5}$.  Assuming the thick disk edge is $\sim$3 kpc \citep[][black dotted lines in the upper-right panel of Figure \ref{fig:zmaxe}]{Carollo2010}, and taking the above observational evidence into account, we conclude that (1) both our N-rich field stars and the control sample stars are mostly halo stars; (2) our N-rich field stars are located closer to the inner halo compared to the control sample. A similar conclusion was also found by \citet{Carollo2013}. 

\begin{figure}
\centering
\includegraphics [width=0.48\textwidth]{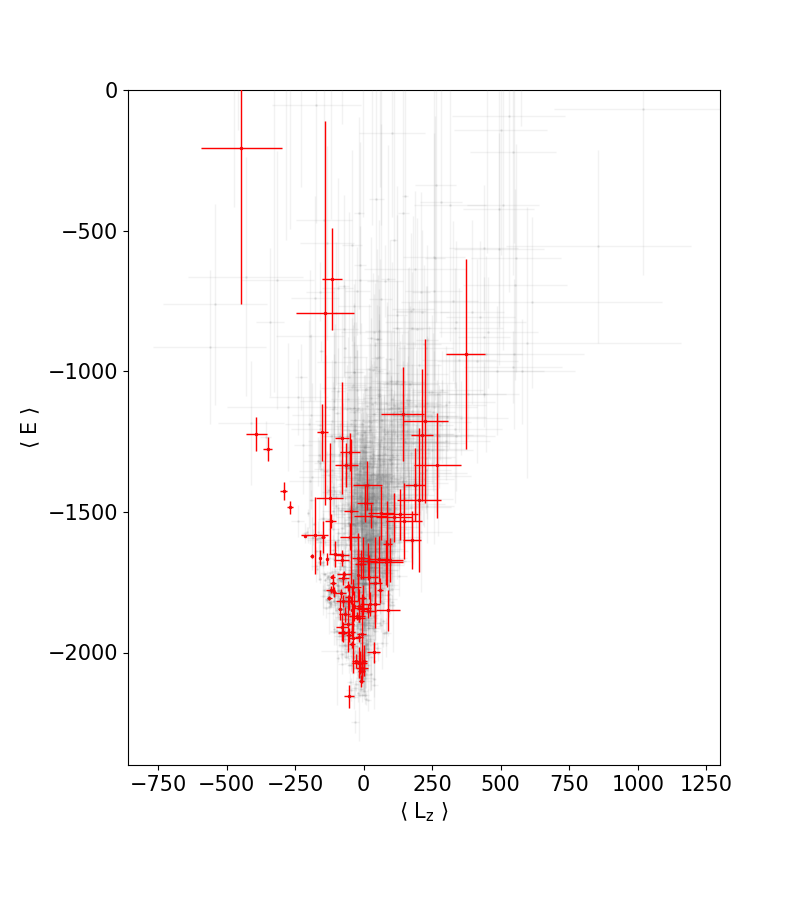} 
\caption{The N-rich field stars (red) and normal metal-poor field stars (gray) in the integral of motion space. 
}\label{fig:ELZ}
\end{figure}

\begin{figure}
\centering
\includegraphics [width=0.48\textwidth]{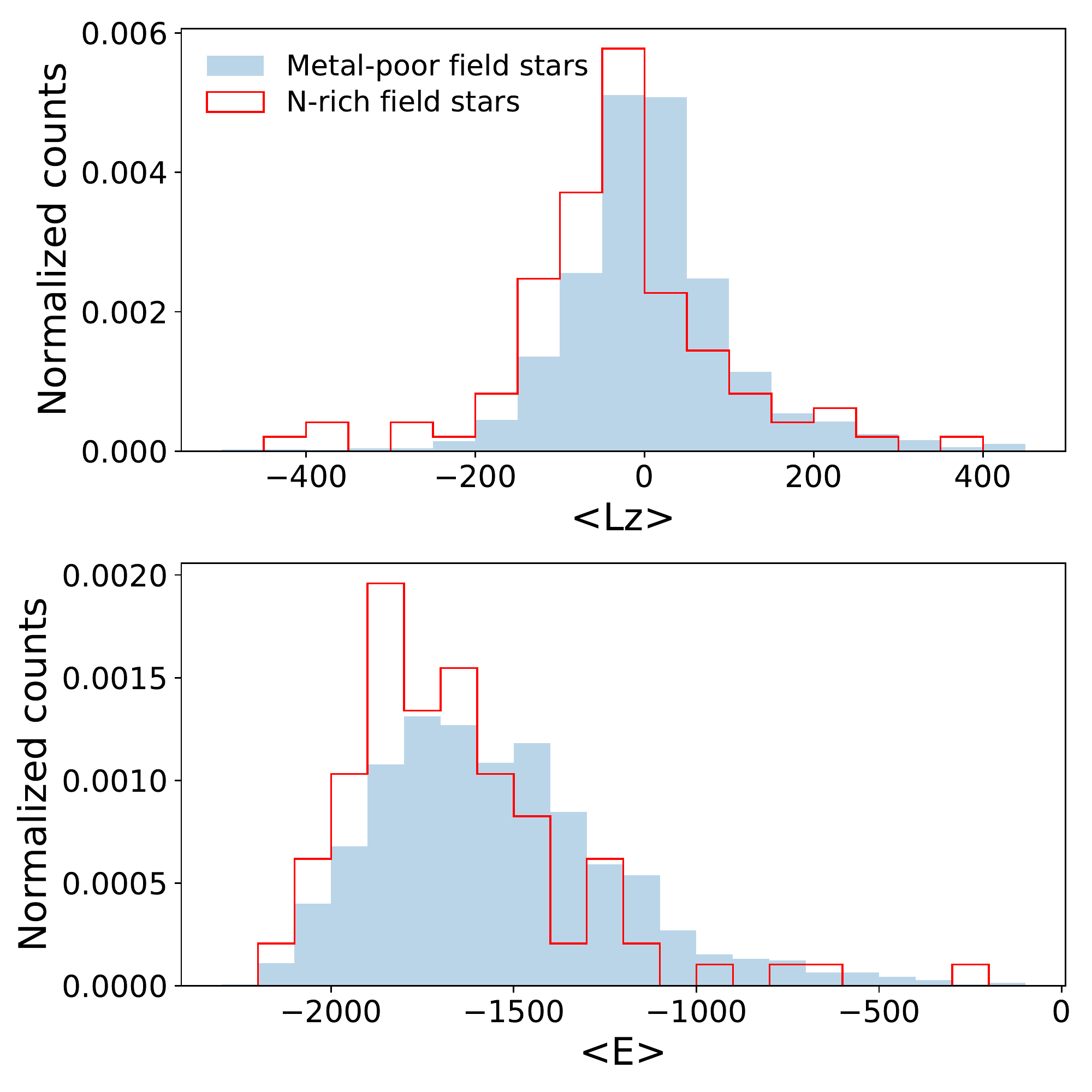} 
\caption{$<E>$ (upper panel) and $<L_{z}>$ (lower panel) distribution histograms of N-rich field stars (red histograms) and normal metal-poor field stars (gray histograms). 
}\label{fig:ELZhist}
\end{figure}

The total orbital energy and z-direction angular momentum of a star should be conserved in a time-invariant gravitational potential. Therefore, the integral of motion space (IoM, E vs. Lz) is a smoking gun to trace the history of a star. Since GravPot16 employs an axis-asymmetric Galactic potential, with a slowly rotating bar in the center of our Galaxy, the total orbital energy and z-direction angular momentum of a star is not constant anymore. In this work, we use the mean of total orbital energy ($<E>$) and the mean of z-direction angular momentum ($<L_{z}>$) over integration time to construct the IoM space. To ensure that we draw solid conclusions from the IoM space, we also simulate the orbits in an axis-symmetric potential. After comparing the results from the two Galactic potential models (e.g., Figure \ref{fig:ELZ}), we find the differences in the IoM space is negligible, mainly because our stars are mostly located in the halo. Thus the following conclusions that we draw from the IoM space are valid in both axis-symmetric and axis-asymmetric Galactic potential models.

Figure \ref{fig:ELZ} shows the IoM space for N-rich field stars and the control sample. The detailed distributions of $<E>$ and $<L_{z}>$ can be found in Figure \ref{fig:ELZhist}. We point out that our $<L_{z}>$ is slightly different than some other studies: positive sign means retrograde, and negative sign means prograde. This is related to the above-mentioned Galactocentric coordinate system that we use in this work. Compared with the locations of different Galactic components in the IoM space (e.g., \citealt{Massari2019}), Figure \ref{fig:ELZ} suggests that except for a small portion of stars located in the thick disk, most of our N-rich field stars are halo stars, which is consistent with what we find using $<Z_{\rm max }>$ (Figure \ref{fig:zmaxe}).  Furthermore, Figure \ref{fig:ELZhist} reveals that N-rich field stars have smaller $<E>$ (less energetic orbits) and $<L_{z}>$ (more prograde orbits) compared with the control sample. The two sample Kolmogorov-Smirnov (K-S) test indicates that the probability that $<E>$ (and  $<L_{z}>$) of the two samples are drawn from the same parent population is lower that $10^{-4}$. The less energetic and more prograde orbits of our N-rich field stars again suggest they are inner halo stars.

\section{Discussion}
\label{sect:dis}

\begin{table*}
\caption{N-rich Field Stars With Possible Li Enrichment.}              
\label{tab:lirich}      
\centering                                      
\begin{tabular}{c c c c c c c c c }         
\hline\hline  
\# &  RA &  DEC & RV (km/s) & Teff (K) & log(g) & [Fe/H]  & A(Li) & Note\\
\hline 
1 &  253.944305  & 21.655846   & -151.44 &4717.13 &1.677& -1.450   &1.160   & Paper II\\ 
2 &  197.656036  & -6.979531    & -12.78 &5173.49 &2.754 &-1.402  & 1.722   & Paper II\\ 
3 &  268.058044  & 26.255920   & -288.09 &5065.61& 2.783& -1.279  & 1.414&Paper II\\ 
4&122.120483   & 1.946907    & 127.52 &4984.40 &2.071& -1.534   &1.382&Paper II\\ 
5&197.834381   & 2.999973    &  22.19 &4955.84& 2.345& -1.245  & 1.414&Paper II\\ 
6&317.852325& -2.385546 &-0.86   &5058.77 &1.871 &-1.069 &1.65    & Paper I\\
7 & 247.988815 &39.067295 &-59.71 & 4988.19 &2.804 &-0.664& 1.518 & Paper I \\
\hline                                             
\end{tabular}

\end{table*}

\subsection{Li abundances}

Li is a volatile element that can be easily destroyed in a high temperature environment. As a star evolves along the RGB, the Li abundance sharply decreases as the star goes through the first dredge-up. After the star reaches the RGB bump, the classical extra-mixing process is suggested to destroy Li, C, and generate N \cite[e.g.,][]{Iben1967,Gratton2000,Charbonnel2007,Charbonnel2010}. But as we have discussed in Paper I and in this paper, the N abundances of our N-rich field stars are clearly larger than the values predicted by classical extra-mixing. Another nuclear process that can produce higher N abundances is needed.

Interestingly, the discovery of Li-rich stars has also defied the classical extra-mixing process, and researchers have modified the classical extra-mixing process to non-traditional ones, e.g.,  enhanced extra-mixing or asymmetric extra-mixing \citep{Yan2018}. The possible discovery of a few N-rich stars in a sample of Li-rich stars (Sbordone, private communication) has inspired us to search for possible Li-rich stars in our sample. Do they have some sort of channel to keep both N and Li at a high level? We determine the Li abundances following the procedure given in \citet{Gao2019}. Briefly speaking, the Li abundance is derived from template matching method to a pure giant sample which contains over 800,000 giant stars from LAMOST data. The templates are synthesized using the SPECTRUM code, based on the stellar parameters provided by the LAMOST pipeline. The intervals of the grid templates are set to be 100K, 0.25 dex, 0.20 dex, and 0.10 dex in Teff, logg, [Fe/H] and Li abundance, respectively. The abundance is determined by fitting a curve to the chi-square array and finding its minimum. Finally, an eye-inspection is used to double-check the matching result, and eliminate the unreliable ones.

Generally speaking, most of the N-rich stars show A(Li)$<1.0$\footnote{A(Li)=log(N$_{\rm Li}$/N$_{\rm H}$)+12, where N$_{\rm Li}$ and N$_{\rm H}$ are the number densities of lithium and hydrogen, respectively.}, suggesting the co-existence of high N and high Li seems to be unlikely. However, we do notice a few possible candidates with higher Li abundances than normal stars (Table \ref{tab:lirich}). Since it is relatively difficult to accurately determine Li abundances for stars with A(Li)$<1.5$ using low-resolution LAMOST spectra, we leave detailed discussion of these interesting stars after obtaining high resolution optical spectra.

\subsection{Galactic vs. extragalactic origin}

\begin{figure}
\centering
\includegraphics [width=0.48\textwidth]{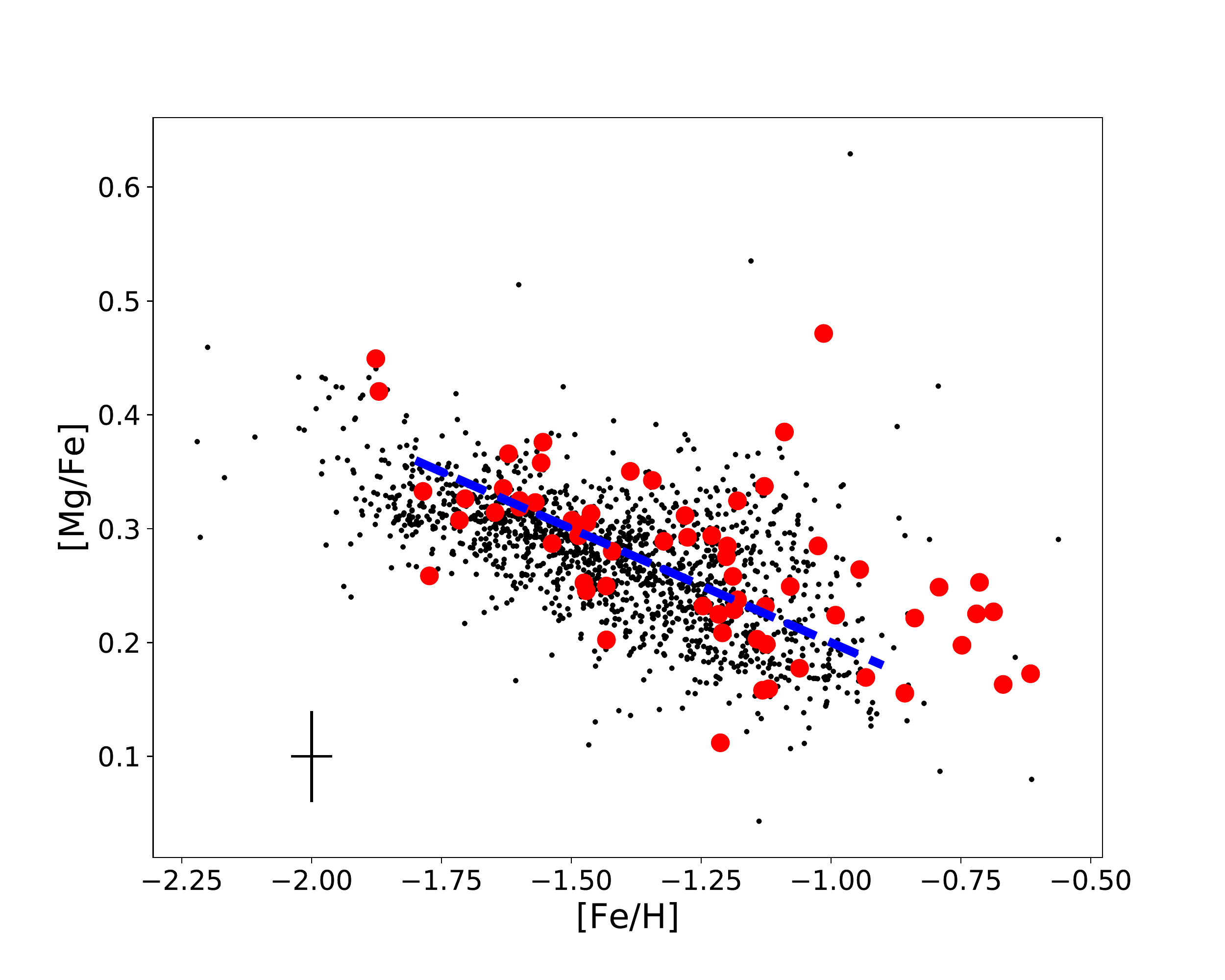} 
\caption{[Mg/Fe] vs. [Fe/H] plot. The N-rich field stars labelled as red dots. The normal metal-poor field stars are shown as black small dots. The line that separates accreted stars and $in-situ$ stars as proposed by \citet{Hayes2018} is labelled as blue dashed line. The typical uncertainties of abundance measurements are shown in the bottom-left as error bars.
}\label{fig:mgfe}
\end{figure}

\begin{figure}
\centering
\includegraphics [width=0.52\textwidth]{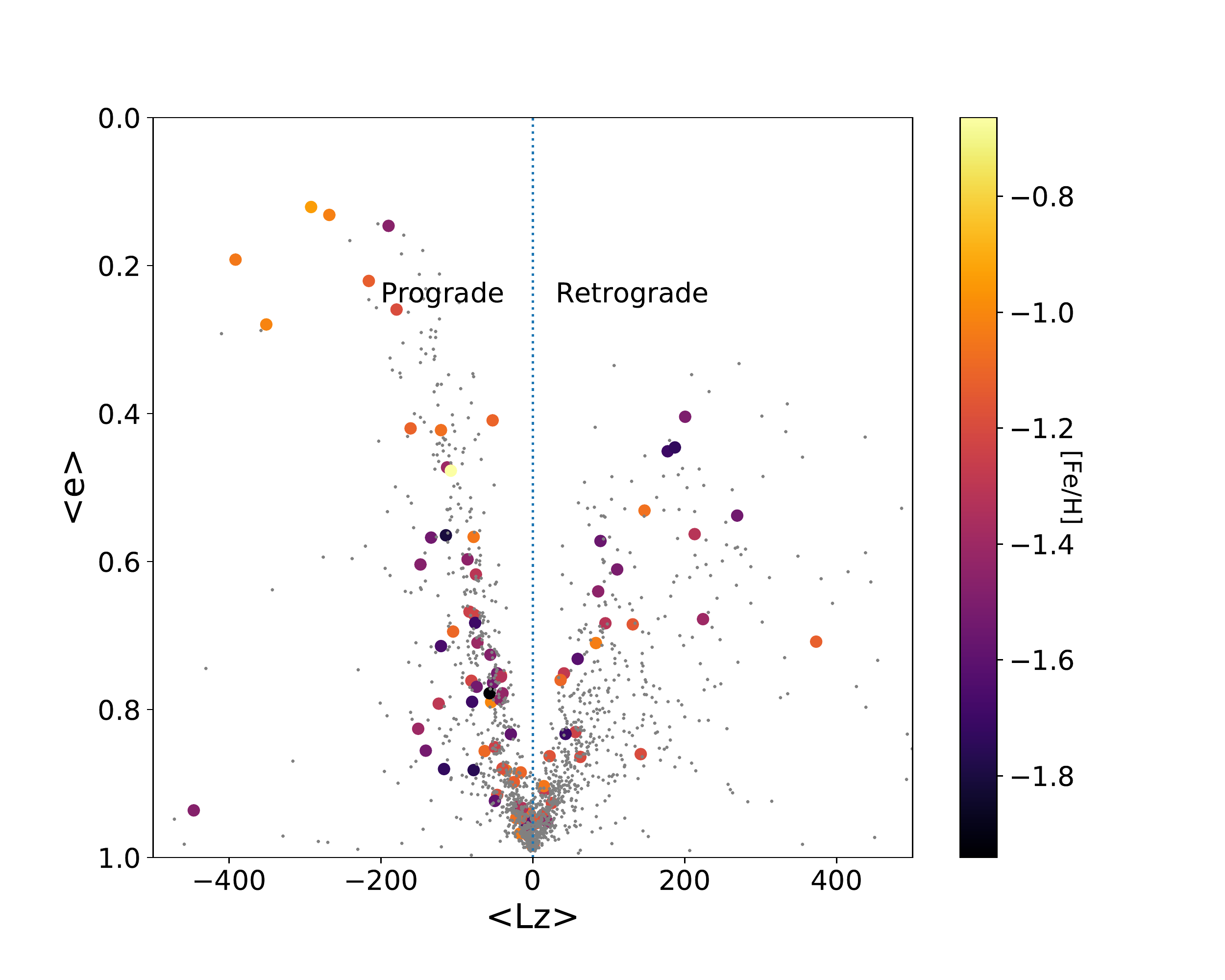} 
\caption{The $<e>$ vs. $<L_{z}>$ plot for N-rich field stars (color dots) and normal metal-poor field stars (gray dots). 
}\label{fig:ecclz}
\end{figure}

Starting from the 90s, astronomers began to apprehend the varieties in the age-metallicity relation and predicted orbits of Galactic GCs (GGCs) \citep{FusiPecci1995, Mackey2004, Forbes2010, Law2010}. GGCs are suggested to be separated into $in-situ$ population and accreted population \citep[e.g.,][]{Forbes2010}. Thanks to the unprecedented accuracy in proper motion and astrometry presented by Gaia, the accreted vs. $in-situ$ theory for GGCs is rapidly developing \citep{Vasiliev2019,Myeong2019,Massari2019}, based on not only the kinematics but also the chemical information from other spectroscopic surveys. The action-angle space and IoM space are suggested to efficiently isolate GCs from different progenitors. In this work, due to the uncertainties in determining distances to halo individual stars, directly linking N-rich field stars to current GGCs with only IoM space distribution (Figure \ref{fig:ELZ}) may over-interpret our data. However, we notice that GGCs are mainly located in the inner halo \citep[e.g., Figure 5 of ][]{Vasiliev2019}, while our N-rich field stars are also mostly inner-halo stars. This generally supports the MW-GC interaction scenario, where inner halo GGCs are prone to stronger dynamical interactions with the MW, and therefore higher possibility to lose (N-rich) stars to the field. In that sense, the Galactic 6D distribution of our bona fide sample of N-rich field stars is a strong observational evidence for galaxy-GC co-evolution simulations.

On the other hand, since a number of GCs are accreted to our MW, if we assume that our N-rich field stars are dissolved from GCs, then we should be able to find both $in-situ$ and accreted stars in our N-rich field star sample.
\citet{Hayes2018} showed that the [Mg/Fe] vs. [Fe/H] may be able to separate stars formed $in-situ$ or $ex-situ$.
Though our LAMOST spectra are low-resolution, where a lot of element lines are blended, one can still derive chemical abundances with uncertainties of $\sim$0.1 dex for the most prominent lines, e.g., Mgb line. Here we use the Mg abundances derived by a data-driven code, SLAM \citep{Zhang2019}, to investigate our N-rich field star sample and control sample.  Figure \ref{fig:mgfe} shows that our N-rich field star sample and the control sample are similarly distributed in the [Mg/Fe] vs. [Fe/H] plane. The line that separates accreted stars and $in-situ$ stars proposed by \citet{Hayes2018} is labelled as blue dashed line in the figure. We do not see a clear separation near the blue dashed line in our data, indicating that two samples consist of both $in-situ$ and accreted stars.
Recently, \citet{Ostdiek2019} complied a catalog of accreted stars with full 6D phase space information, and extended the catalog to stars with only 5D information using the machine learning technique. 
We find 11 common stars between the 6D phase space accreted star catalog and our N-rich field stars from Paper I and this paper. Four stars are labelled as accreted stars. With the caution of small statistics (4 out of 11 stars), it seems that a substantial portion of our N-rich field stars may be accreted to the MW.  The above-mentioned evidences seem to further strengthen the GC origin scenario of the N-rich field stars.

\citet{Savino2019} compared GGCs and a population of CN-strong stars in terms of kinematics, e.g., IoM space distributions. They suggested that a group of low-circularity stars with [Fe/H$]\sim -1$ may come from the outer Galactic disk.  Here we use eccentricity instead of circularity, given that high-circularity stars tend to have low eccentricity. We directly plot eccentricity as a function of $<L_{z}>$ (using [Fe/H] as color) in Figure \ref{fig:ecclz}. Low eccentricity stars tend to have more circular orbits and higher angular momentum, for both the N-rich field star sample and the control sample. This trend seems to be continuous. The clustering at [Fe/H$]\sim -1$ for low eccentricity stars as suggested by \citet{Savino2019} is vaguely seen ($\sim 11$ stars) in our work. If such group of N-rich field stars with disk-like kinematics and  [Fe/H$]\sim -1$ do exists, their GC origin scenario may still hold, since there are evidences that support the existence of thick disk GCs, e.g., Figure 4 of \citet{Massari2019} and NGC 5927 \citep{Allen2008, Mura2018}.

If N-rich field stars are originated from existing/dissolved GCs, then a bona fide sample of N-rich field stars would be crucial to understand the formation and co-evolution of MW and GCs (e.g., E-MOSAIC, \citealt{Kruijssen2019}). 

\subsection{AGB contaminated materials}
\label{sect:AGB}

Recent discovery of a N-rich, mildly metal-poor ([Fe/H$]=-1.08$) giant star in a single-lined spectroscopic system by \citet{FT2019binary} has inspired the scenario that some of the N-rich metal-poor ``field'' stars may reside in binary systems, where the AGB companion stars have died out. 
A binary AGB companion scenario was also invoked by \citet{Simpson2019} to explain a N-rich ([N/Fe$]>+2.5$), metal-poor ([Fe/H$]<-2$) star found in the globular cluster ESO280-SC06. According to nucleosynthetic theories \citep[e.g.,][]{Masseron2010, Karakas2014}, high N abundances can be seen in intermediate-mass AGB stars, where Hot-Bottom Burning (HBB) can produce N at the expense of C.

On the other hand, several theories have been proposed to explain the chemical enriched populations in GCs, e.g., AGB-ejecta \citep[e.g.,][]{DErcole2008,DErcole2010,Ventura2013}, fast-rotating massive stars \citep{Decressin2007}, massive binaries \citep{deMink2009}, supermassive stars \citep{Denissenkov2014}, etc. However, none of them seem to explain all the observational evidences of GCs \citep[e.g.,][]{Bastian2018}. In spite of its need for improvements, the AGB-ejecta scenario stands out for its ability to explain the observed N, Na(, Al) enhancement and C, O(, Mg) depletion in SG stars. In that scenario, the chemical peculiarity is attributed to the HBB phase nucleosynthesis of AGB stars.  Recently, \citet{Bekki2019} proposed that the high-density building blocks of the Galactic bulge may also generate suitable AGB ejecta, which can explain the N-enhancement of some bulge field stars.

There is one thing in common among these scenarios: AGB ejecta.
Therefore, we may only be able to confirm the importance of AGB ejecta in N-rich field stars from the chemical point of view. In that sense, the kinematic information is the key to disentangle these three scenarios. To verify the binary AGB companion scenario demands further observational data, where light curves and RV changes are indicators of the presence of binary systems. However, the inner-halo-like orbits of our N-rich field stars seem to favor the GC scenario over the bulge scenario (Section \ref{sect:orb}).

\section{Conclusion}
\label{sect:con}

Since the discovery of N-rich field stars, astronomers have proposed different scenarios to explain this phenomenon, including GC, AGB binary, Galactic bulge, and etc. A comprehensive bona fide sample with detailed chemical and kinematic information is needed to unveil the truth. In this paper, we have extended our search for N-rich field stars to LAMOST DR5, 
where we efficiently identified $\sim 100$ such stars with CN-CH bands. We first investigate chemical abundances through seven common stars with APOGEE high resolution spectra. The Mg, Al, and Si abundances of these common stars generally agree with that of GC enriched stars, but it is still inconclusive for C, N, and O. On the other hand,  the orbits of N-rich field stars show similar properties as inner-halo stars: lower $<Z_{\rm max}>$, $<r_{apo}>$, $<L_{z}>$, and $<E>$ compared to normal metal-poor halo field stars. The kinematics of N-rich field stars and GC seem to share similarities. The lack of multi-epoch RV data prevent us from drawing conclusion on binary AGB companion scenario. We are aware of the limited sample size of N-rich field stars with high-resolution spectra, from where we derived chemical abundances.
To achieve a statistically significant conclusion, we plan to obtain high-resolution spectra for 20$-$30 N-rich field stars. 
Detailed chemical patterns and multi-epoch RV data of a statistically significant sample of N-rich field stars will further constrain their formation scenarios. 


\section{acknowledgments}
We thank Luca Sbordone, Yue Wang and Bryan Ostdiek for helpful discussions. We thank the anonymous referee for insightful comments. 
Baitian Tang gratefully acknowledges support from National Natural Science Foundation of China under grant No. U1931102 and support from the hundred-talent project of Sun Yat-sen University. J.G.F-T is supported by FONDECYT No. 3180210 and Becas Iberoam\'erica Investigador 2019, Banco Santander Chile.
This research is supported by National Natural Science Foundation of China under grant Nos. 11603037 and 11973052. H.-L.Y. acknowledges supports from Youth Innovation Promotion Association, CAS, and the supports from the Astronomical Big Data Joint Research Center, co-founded by the National Astronomical Observatories. D.G. gratefully acknowledges support from the Chilean Centro de Excelencia en Astrof\'isica
y Tecnolog\'ias Afines (CATA) BASAL grant AFB-170002.
D.G. also acknowledges financial support from the Direcci\'on de Investigaci\'on y Desarrollo de
la Universidad de La Serena through the Programa de Incentivo a la Investigaci\'on de
Acad\'emicos (PIA-DIDULS).

Guoshoujing Telescope (the Large Sky Area Multi-Object Fiber Spectroscopic Telescope LAMOST) is a National Major Scientific Project built by the Chinese Academy of Sciences. Funding for the project has been provided by the National Development and Reform Commission. LAMOST is operated and managed by the National Astronomical Observatories, Chinese Academy of Sciences. Funding for the GravPot16 software has been provided by the Centre national d’etudes spatiale (CNES) through grant 0101973 and UTINAM Institute of the Université de Franche-Comté supported by the Région de Franche-Comté and Institut des Sciences de l’Univers (INSU). Monte Carlo simulations have been executed on computers from the Utinam Institute of the Université de Franche-Comté supported by the Région de Franche-Comté and Institut des Sciences de l’Univers (INSU).

\bibliographystyle{aasjournal}
\bibliography{gc,survey,cnchcor}

\label{lastpage}

\end{document}